\documentclass[10pt,final,journal,twocolumn]{IEEEtran}
\ifCLASSINFOpdf

\else
\usepackage{algorithm}
\usepackage{booktabs}

\fi
\usepackage{cite}
\usepackage[cmex10]{amsmath}
\usepackage{amsthm}
\usepackage{stfloats}
\usepackage{multicol,multienum}
\usepackage{multirow}

\usepackage{amssymb}
\usepackage{url}
\usepackage{amsmath}
\usepackage{xcolor}
\usepackage[dvips]{graphicx}
\usepackage{subfigure}
\usepackage{caption}
\usepackage{amsmath}
\usepackage{amsfonts,amssymb}
\usepackage{bbm}
\usepackage[T1]{fontenc}
\usepackage{algorithm}
\usepackage{algpseudocode}
\usepackage{ulem}
\usepackage{colortbl}  %彩色表格需要加载的宏包
\usepackage{xcolor}
\usepackage[linesnumbered, ruled]{}
\makeatletter
\renewcommand{\maketag@@@}[1]{\hbox{\m@th\normalsize\normalfont#1}}%
\makeatother

\begin{document}
% correct bad hyphenation here
\hyphenation{op-tical net-works semi-conduc-tor}
%\begin{document}

\title{Movable Antennas-Assisted Secure Transmission Without Eavesdroppers' Instantaneous CSI}
\author{Guojie Hu, Qingqing Wu,~\textit{Senior Member}, \textit{IEEE}, Donghui Xu, Kui Xu,~\textit{Member}, \textit{IEEE}, Jiangbo Si,~\textit{Senior Member}, \textit{IEEE}, Yunlong Cai,~\textit{Senior Member}, \textit{IEEE} and Naofal Al-Dhahir,~\textit{Fellow}, \textit{IEEE}%\vspace{-2.25em}
 %and Yunlong Cai,~\textit{Senior Member}, \textit{IEEE}
\thanks{
%This work was supported in part by the Natural Science Foundations of China under Grants 62071480 and 61971337 and in part by the National Natural Science Foundation
%for Distinguished Young Scholar under Grant 61825104. (Corresponding author: Jiangbo Si)
%This work was supported in part by the Natural Science Foundations
%of China under Grants 62201606.
Guojie Hu and Donghui Xu are with the College of Communication Engineering, Rocket Force University of Engineering, Xi'an 710025, China (email: lgdxhgj@sina.com; xdh\_45@163.com). Qingqing Wu is with the Department of Electronic Engineering, Shanghai Jiao Tong University, Shanghai 200240, China (qingqingwu@sjtu.edu.cn). Kui Xu is with the College of Communications Engineering, Army Engineering University of PLA, Nanjing 210007, China (lgdxxukui@sina.com). Jiangbo Si is with the Integrated Service Networks Lab of Xidian University, Xi'an 710100, China (jbsi@xidian.edu.cn). Yunlong Cai is with the College of Information Science and Electronic Engineering, Zhejiang University, Hangzhou 310027, China (email: ylcai@zju.edu.cn). Naofal Al-Dhahir is with the Department of Electrical and Computer Engineering, The University of Texas at Dallas, Richardson, TX 75080 USA (aldhahir@utdallas.edu).
}\vspace{-0.001em}
}
%This work was supported by the Natural Science Foundations of China (No. 61671474).
%L. X. Yang, D. Wu, and Y. M. Cai are with the College of Communications Engineering, the Army of Engineering University, Nanjing 210007, China. (Email: yanglianxin1228@126.com; wujing1958725@126.com; caiym@vip.sina.com.
\IEEEpeerreviewmaketitle
%\vspace{-5pt}
\maketitle
%\vspace{-30pt}
\begin{abstract}
Movable antenna (MA) technology is highly promising for improving communication performance, due to its advantage of flexibly adjusting positions of antennas to reconfigure channel conditions. In this paper, we investigate MAs-assisted secure transmission under a legitimate transmitter Alice, a legitimate receiver Bob and multiple eavesdroppers. Specifically, we consider a practical scenario where Alice has no any knowledge about the instantaneous non-line-of-sight component of the wiretap channel. Under this setup, we evaluate the secrecy performance by adopting the secrecy outage probability metric, the tight approximation of which is first derived by interpreting the Rician fading as a special case of Nakagami fading and concurrently exploiting the Laguerre series approximation. Then, we minimize the secrecy outage probability by jointly optimizing the transmit beamforming and positions of antennas at Alice. However, the problem is highly non-convex because the objective includes the complex incomplete gamma function. To tackle this challenge, we, for the first time, effectively approximate the inverse of the incomplete gamma function as a simple linear model. Based on this approximation, we arrive at a simplified problem with a clear structure, which can be solved via the developed alternating projected gradient ascent (APGA) algorithm. Considering the high complexity of the APGA, we further design another scheme where the zero-forcing based beamforming is adopted by Alice, and then we transform the problem into minimizing a simple function which is only related to positions of antennas at Alice. Such problem is well-solved via another projected gradient descent algorithm developed with a lower complexity. As demonstrated by simulations, our proposed schemes achieve significant performance gains compared to conventional schemes based on fixed-position antennas.

\end{abstract}
%\vspace{-15pt}

\IEEEpeerreviewmaketitle
%\vspace{-15pt}
\section{Introduction}
\IEEEPARstart{W}{ith} the explosive growth of data traffic in 5G/B5G and future 6G, there will be a higher risk that a large amount of private data can be overheard by malicious eavesdroppers \cite{huangG}. Physical-layer security (PLS), by exploiting the inherent properties of wireless channels (for instance, random fading and noise), is expected to realize substantial and fundamental secrecy for wireless communication, regardless of how powerful the eavesdroppers' computing ability is \cite{JS_IT}.

To improve the secrecy performance of PLS, numerous technologies have been proposed, such as secure beamforming \cite{globcom}, cooperative jamming \cite{caokunrui_TIFS}, artifical noise \cite{fangsaisai_CL} and channel-based secret key \cite{WCM}, among which secure beamforming always plays an important role because it effectively improves/reduces the receiving power gains at the legitimate receiver/malicious eavesdroppers by carefully designing antenna weights.

Conventional secure beamforming schemes all rely on fixed-position antennas (FPAs), where positions of antennas at the transmitter/receiver cannot be changed \cite{Xiaozhenyu}. This limitation of FPAs will result in a fundamental deficiency, i.e., when the direction angles of the eavesdroppers are close to that of the legitimate receiver, the correlation between the main channel and the wiretap channel will become much higher. Then, beamforming with FPAs cannot effectively distinguish between the legitimate receiver and the eavesdroppers, and thus their signal receiving quality will be similar, leading to a poor secrecy performance \cite{SPL}.

Recently, researchers have proposed a novel technology of movable antennas (MAs) \cite{Zhulipeng_CM} or fluid antennas \cite{FA1,FA_CL,FA_MA}. Specifically, employing MAs, antennas (at the transmitter or the receiver) are connected to radio frequency (RF) chains via flexible cables, and positions of antennas can be flexibly adjusted in the specified region supported by the stepper motors or servos. This feature of MAs allows for real-time reconfiguration of wireless channels and thus brings an additional spatial degree of freedom (DoF). Therefore, using MAs, even when the direction angles of the eavesdroppers and the legitimate receiver are similar, their channel correlation can be proactively reduced by optimizing antennas' positions and then beamforming can unlock its true advantages \cite{Zhulipeng_CL,guojie_irs}. Further, compared to other schemes such as i) antenna selection \cite{Fast_TSP}, which requires more candidate antennas, RF chains, expensive hardware cost and complex channel estimation, and ii) rotatable uniform linear array \cite{RULA_TWC}, which only mechanically rotates the transmit/receive array and cannot exploit spatial channel variation, MAs fully utilize the channel variation in the continuous spatial region to achieve a higher spatial diversity without no excessive demand for additional hardware costs or algorithm overheads.

Motivated by these potential merits, early works have considered integrating MAs into various system setups \cite{Zhulipeng_CM,Flexible_MIMO_WCM,guojie_irs,qinMA,Zhulipeng_modelling,Mawenyan_compressed,Xu_hao_channel_estimation,Dailinglong_arxiv}. For instance, \cite{Zhulipeng_CM} and \cite{Flexible_MIMO_WCM} first provided an overview of the promising applications and hardware designs for MAs-aided communications, and concurrently the advantages of employing MAs for signal power improvement, interference mitigation, flexible beamforming and spatial multiplexing. Subsequently, \cite{Zhulipeng_modelling} proposed the field-response-based channel model, and provided theoretical analysis for the maximum channel gain achieved by a single MA in both deterministic and stochastic channels. In \cite{Mawenyan_compressed,Xu_hao_channel_estimation,Dailinglong_arxiv}, compressed sensing, low-sample-size and successive Bayesian reconstructor based channel estimations for MAs-aided systems were presented. In \cite{Mawenyan_MIMO,Gao_Xiqi_CL,Yongpeng_WU_Glob}, MAs-aided multiple-input multiple-output (MIMO) system was investigated, where in addition to the covariance of the transmit signals, positions of MAs at the transmitter/receiver are also carefully optimized to enhance the capacity. In \cite{Pixiangyu_GC,Guojie_Fluid,Songjie_WCL,Yifei_Wu_arxiv}, MAs-enabled multiuser uplink/downlink was studied, which demonstrated that the total transmit power (sum rate) of all users can be further reduced (improved) by additionally optimizing positions of MA(s) at each user or the base station. Recently, \cite{Xiaodan_Shao_arxiv,Xiaodan_Shao_arxiv1} further proposed a novel six-dimensional movable antenna (6DMA) architecture, where 3D positions and 3D rotations of antenna surfaces can be concurrently optimized to fully exploit the spatial DoF for achieving the greatest flexibility and then the highest capacity improvement. Besides, MAs-enabled wideband communications, interference networks, symbiotic radio communications, coordinated multi-point (CoMP) reception systems and over-the-air computation were further considered in \cite{Zhulipeng_arxiv1,Wanghonghao_arxiv1,Lyubin_arxiv1,Guojie_COMP,fa_air_computing}. In addition to these works, exploiting MAs for enhancing the secrecy performance has also been studied in \cite{Chenzhenqiao_ICASSP,Guojie_SPL,Pancunhua_arxiv}. In detail, under the setups of multiple-input single-output single/multiple eavesdropper(s) (MISOSE/MISOME), \cite{Chenzhenqiao_ICASSP,Guojie_SPL} considered the problem of maximizing the secrecy rate and minimizing the power consumption via the optimizations of MAs' positions and beamforming at the transmitter. Under the setup of MIMOME, \cite{Pancunhua_arxiv} intended to maximize the secrecy rate by jointly optimizing the transmit precoding matrix, the artificial noise covariance matrix and MAs' positions.

In this paper, we study MAs-assisted secure transmission under the setup of MISOME, where the private information is transmitted from the MAs-enabled Alice to the single-antenna Bob, in the presence of multiple single-antenna and colluding eavesdroppers. Unlike previously related works \cite{Chenzhenqiao_ICASSP,Guojie_SPL,Pancunhua_arxiv}, we consider the general Rician fading for the wiretap channel, for which Alice does not know its instantaneous non-line-of-sight (NLoS) component and just masters its statistical LoS component via channel environment measures, global positioning systems and so on \cite{Wang_huiming_SPL,IOT_Wuhuici,Libin_TWC,RIS_outage_probability}. Due to the randomness of the NLoS component of the wiretap channel, the instantaneous secrecy rate cannot be obtained anymore. Therefore, unlike \cite{Chenzhenqiao_ICASSP,Guojie_SPL,Pancunhua_arxiv}, we adopt the secrecy outage probability as the secrecy performance evaluation metric and aim to reveal the advantages of MAs to improve the secrecy performance, even with relatively statistical channel state information. Specifically, the main contributions of this paper are summarized as follows.
 \begin{itemize}
 \item We consider a pessimistic eavesdropping scenario where multiple eavesdroppers aim to jointly decode the private information of Alice. This behavior makes it very hard to derive the exact secrecy outage probability expression. To overcome this challenge, we first regard the Rician fading as a special case of Nakagami fading and concurrently exploit the Laguerre series approximation to obtain a tight and closed-form secrecy outage probability, based on which numerous insights are gained and discussed.
      \end{itemize}
 \begin{itemize}
 \item We minimize the secrecy outage probability by jointly optimizing the transmit beamforming and positions of antennas at Alice. However, the problem is highly non-convex since the objective includes the complex incomplete gamma function, and more importantly, the shape and scale parameters of which all involve the optimization variables. Facing this challenge, we introduce the slack variable and place the original objective in the constraint. Afterwards, we propose a novel linear model to effectively approximate the inverse of the incomplete gamma function, based on which the problem can be successfully transformed into a simpler one with a clear structure.
      \end{itemize}
       \begin{itemize}
 \item The simplified problem, although more concise, is still highly non-convex. Addressing this challenge, we develop a general alternating projected gradient ascent (APGA) algorithm to iteratively optimize the transmit beamforming and positions of antennas at Alice to obtain a stationary solution. Furthermore, considering the higher complexity of the APGA, we also propose a sub-optimal scheme where the zero-forcing (ZF)-based transmit beamforming is adopted at Alice to completely eliminate the averaged receiving power at eavesdroppers via the LoS paths, according to which the problem can be simplified into minimizing a simple function which is only related to positions of antennas at Alice. Then, another projected gradient descent (PGD) algorithm is further designed to effectively solve the problem with a lower complexity.
  \item We conduct numerical simulations and provide several competitive benchmarks to show the effectiveness of the proposed schemes. We reveal that, i) compared to the conventional FPAs, MAs with careful optimizations of antennas' positions can achieve a significant performance gain by exploiting the additional spatial DoF; ii) the ZF-based transmit beamforming works well when the difference in direction angles between the main channel and the wiretap channel is higher; iii) the secrecy outage probability decreases and then converges to a constant with respect to (w.r.t.) the length of antennas' movement region, indicating that only a limited span is enough to achieve the optimal performance.
 \end{itemize}
 The rest of this paper is organized as follows. The system model and problem formulation are presented in Sections II. The secrecy outage probability analysis and some insights are detailed in Section III. The proposed novel optimization framework for minimizing the secrecy outage probability is provided in Section IV, and another competitive benchmark based on the ZF-based transmit beamforming is given in Section V. The simulations are presented in Section VI and the conclusions are shown in Section VII.
 \newcounter{mytempeqncnt}
 \begin{figure}
% \vspace{-10pt}
\centering
\includegraphics[width=8cm]{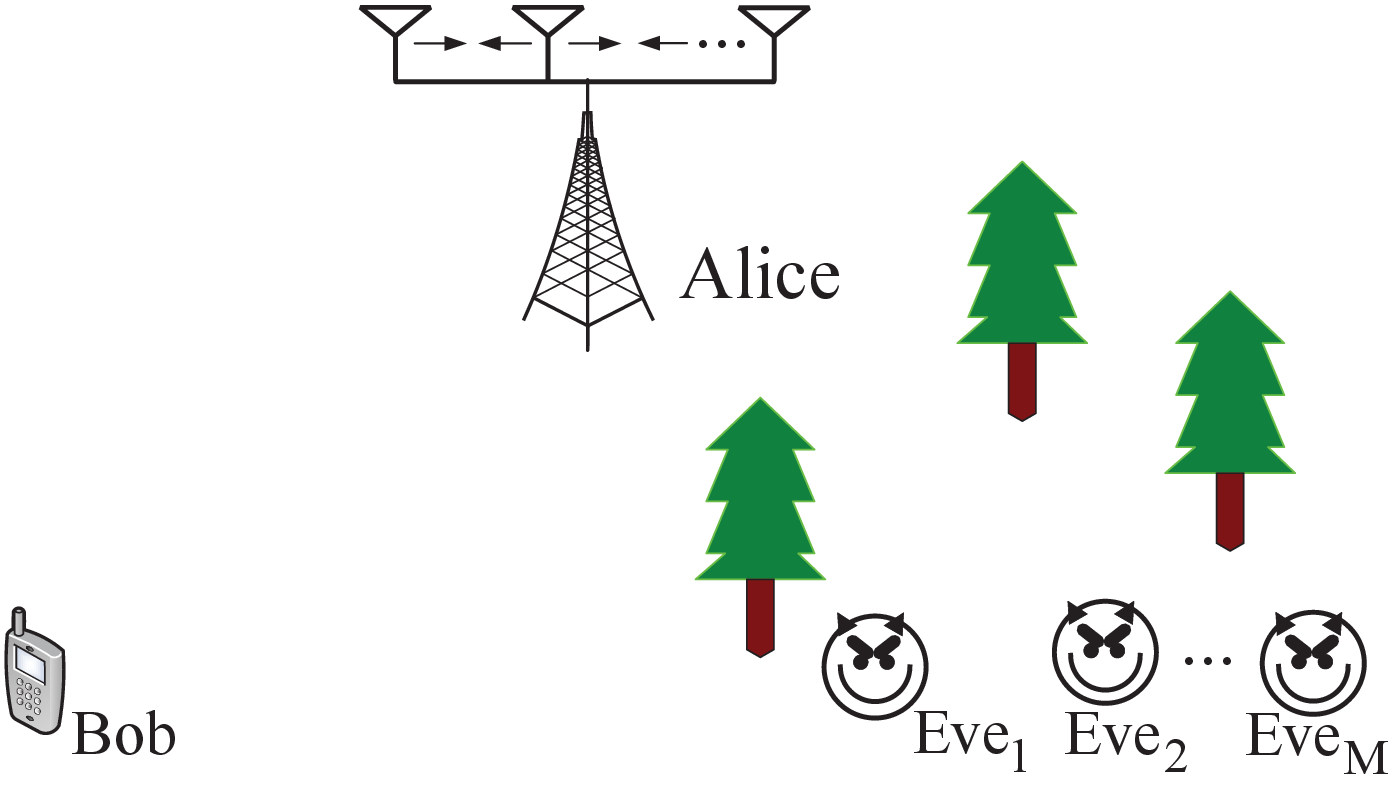}
\caption{{System model.}} \label{fig:Fig1}
\vspace{-15pt}
\end{figure}
\vspace{-5pt}
\section{System Model and Problem Formulation}
%\subsection{System Model}
%\newcounter{mytempeqncnt}
\subsection{System Model}
As illustrated in Fig. 1, this paper considers a downlink secure transmission system consisting of a transmitter (Alice), a legitimate receiver (Bob), and $M \ge 2$ eavesdroppers, denoted as $\left\{ {{\rm{Ev}}{{\rm{e}}_i}} \right\}_{i = 1}^M$. Bob and $\left\{ {{\rm{Ev}}{{\rm{e}}_i}} \right\}_{i = 1}^M$ are equipped with a single antenna each, while Alice is equipped with a linear MA array consisting of $N \ge 2$ antennas.\footnotemark \footnotetext{Note that Alice can also adopt the planar array. Nevertheless, all conclusions obtained in this paper are not affected by this extension. Therefore, we consider a simple setting for a clear presentation.} Since the linear array is assumed, we denote ${\left[ {{x_1},{x_2},...,{x_N}} \right]^T} \buildrel \Delta \over = {\bf{x}}$ as the one-dimensional positions of $N$ MAs relative to the reference point zero, where ${\left(  \cdot  \right)^T}$ is the transpose operation. Without loss of generality, we set $0 \le {x_1} < {x_2} < ... < {x_N} \le L$, with $L$ denoting the total span for the movement of MAs. Note that unlike conventional FPAs, positions of these $N$ MAs, i.e., $\left\{ {{x_i}} \right\}_{i = 1}^N$, can be flexibly adjusted in the specific region to proactively cater to effective beamforming designs, such that the secrecy performance can be further improved. The details are shown later.

The $1 \times N$ main channel vector and the $i$-th wiretap channel vector from Alice to Bob and from Alice to ${\rm{Eve}}_i$, are respectively denoted as ${{\bf{h}}_0}({\bf{x}})$ and ${{\bf{h}}_i} ({\bf{x}})$, $i = 1,2,...,M$. Unlike \cite{Chenzhenqiao_ICASSP,Guojie_SPL,Pancunhua_arxiv}, we in this paper adopt the general Rician fading model for the wiretap channel, because eavesdroppers may deliberately hide by the trees and other structures to avoid being discovered and then there may exist not only the LoS channel path, but also the NLoS channel paths. Therefore, ${{\bf{h}}_i}({\bf{x}})$, $\forall i = 1,...,M$, has the following form
%{\setlength\abovedisplayskip{1pt}
%\setlength\belowdisplayskip{1pt}
\begin{equation}
{{\bf{h}}_i}({\bf{x}}) = \sqrt {\frac{{{K_i}{\beta _i}}}{{{K_i} + 1}}} {\overline {\bf{h}} _i}({\bf{x}}) + \sqrt {\frac{{{\beta _i}}}{{{K_i} + 1}}} {\widetilde {\bf{h}}_i},
 \end{equation}
where $K_{i} > 0$ is the Rician K-factor, ${{\beta _{i}}}$ is the path loss component, ${\widetilde {\bf{h}}_{i}} \in {{\mathbb{C}}^{1 \times N}}$ is the NLoS component, and entries of ${\widetilde {\bf{h}}_{i}}$ are independent and identically
distributed (i.i.d) circularly-symmetric complex Gaussian random
variables with zero mean and unit variance, i.e., ${\widetilde {\bf{h}}_{i}} \sim {\cal C}{\cal N}\left( {0,{{\bf{I}}_N}} \right)$. In addition, ${\overline {\bf{h}} _{i}}({\bf{x}})$ is the LoS component, which is related to positions of Alice's antennas ${\bf{x}}$ and can be expressed as\footnotemark \footnotetext{Actually, the instantaneous ${\widetilde {\bf{h}}_i}$ corresponding to different ${\bf{x}}$ may be also different. However, as explained later, we in this paper consider a practical limitation that Alice only masters the statistical information of ${\widetilde {\bf{h}}_i}$. Therefore, from the perspective of Alice, the statistical distribution of ${\widetilde {\bf{h}}_i}$ is the same for different ${\bf{x}}$.}
\begin{equation}
{\overline {\bf{h}} _i}({\bf{x}}) = \left[ {{e^{j\frac{{2\pi }}{\lambda }{x_1}\sin {\theta _i}}},{e^{j\frac{{2\pi }}{\lambda }{x_2}\sin {\theta _i}}},...,{e^{j\frac{{2\pi }}{\lambda }{x_N}\sin {\theta _i}}}} \right],
 \end{equation}
where ${{\theta _i}}$ is the angle of departure (AoD) from the linear array at Alice to ${\rm{Eve}}_i$, and $\lambda $ is the carrier wavelength. Without loss of generality, it is assumed that ${\theta _i} \ne {\theta _j}$, $\forall i,j \in \left\{ {1,2,...,M} \right\}$ and $i \ne j$.

On the other hand, unlike eavesdroppers, Bob can communicate with Alice without any need for concealment. Hence, for simplification we consider the LoS condition for the main channel. Therefore, ${{{\bf{h}}_0}({\bf{x}})}$ is expressed as
\begin{equation}
{{\bf{h}}_0}({\bf{x}}) = \sqrt {{\beta _0}} \left[ {{e^{j\frac{{2\pi }}{\lambda }{x_1}\sin {\theta _0}}},{e^{j\frac{{2\pi }}{\lambda }{x_2}\sin {\theta _0}}},...,{e^{j\frac{{2\pi }}{\lambda }{x_N}\sin {\theta _0}}}} \right],
 \end{equation}
where ${{\beta _0}}$ is the path loss component and ${{\theta _0}}$ is the AoD from Alice to Bob.\footnotemark \footnotetext{Similar to \cite{Zhulipeng_CL}, we assume the far-field condition between Alice and Bob (eavesdroppers) because the size of the moving region for all antennas is much smaller than the signal propagation distance. Then, the AoDs do not change for different positions of antennas.}

Based on the above analysis, the received signal at Bob is expressed as
\begin{equation}
{y_0} = \sqrt {{P_a}} {{\bf{h}}_0}({\bf{x}}){{\bf{w}}}{x_s} + {n_0},
 \end{equation}
 where ${\bf{w}} \in {{\mathbb{C}}^{N \times 1}}$ is the transmit beamforming vector at Alice, with $\left\| {\bf{w}} \right\| = 1$, $x_s$ is the Gaussian distributed information bearing signal satisfying ${\mathbb{E}}\left[ {{{\left| {x_s} \right|}^2}} \right] = 1$, $n_0$ is the additive white Gaussian noise with zero mean and variable ${\sigma ^2}$, and ${{P_a}}$ is Alice's power budget. Next, we will present the detailed proof that why Alice should set its transmit power to the power budget value.

  Based on (4), the receiving signal-to-noise ratio (SNR) at Bob is derived as
 \begin{equation}
{\gamma _0}({\bf{w}},{\bf{x}}) = {P_a}{\left| {{{\bf{h}}_0}({\bf{x}}){\bf{w}}} \right|^2}/{\sigma ^2}.
 \end{equation}

 Also, we can use the similar manner to derive the receiving SNR at ${\rm{Eve}}_i$ as
  \begin{equation}
 {\gamma _i}({\bf{w}},{\bf{x}}) = {P_a}{\left| {{{\bf{h}}_i}({\bf{x}}){\bf{w}}} \right|^2}/{\sigma ^2}.
 \end{equation}

 We consider a pessimistic eavesdropping scenario where $M$ colluding eavesdroppers can perform joint decoding for $x_s$. Hence, the effective receiving SNR of the wiretap channel is expressed as
 \begin{equation}
 {\gamma _e}({\bf{w}},{\bf{x}}) = {P_a}\sum\nolimits_{i = 1}^M {{{\left| {{{\bf{h}}_i}({\bf{x}}){\bf{w}}} \right|}^2}/{\sigma ^2}}.
 \end{equation}

\subsection{Secrecy Rate and Secrecy Outage Probability}
The secrecy rate in PLS is denoted as ${C_s} = {({C_m} - {C_e})^ + }$, where ${(x)^ + } \buildrel \Delta \over = \max (x,0)$, ${C_m}$ and ${C_e}$ are the capacities of the main channel and the wiretap channel, respectively. Based on the above analysis, $C_m$ and $C_e$ are given by
\begin{equation}
\begin{split}{}
{C_m} =& {\log _2}(1 + {\gamma _0}({\bf{w}},{\bf{x}})),\\
{C_e} =& {\log _2}(1 + {\gamma _e}({\bf{w}},{\bf{x}})).
\end{split}
 \end{equation}

Generally, it is practical for Alice to obtain the CSI of the main channel. Therefore, the rate of transmitted codewords at Alice can be adaptively set as $C_m$. On the other hand, consider the challenging case where eavesdroppers are passive and do not emit any signals, such that Alice has no any knowledge about the random components of the wiretap channel, i.e., ${\widetilde {\bf{h}}_i}$, $i = 1,2,...,M$, but can only estimate the statistical components of the wiretap channel, i.e., $K_i$, ${{\beta _i}}$ and ${\overline {\bf{h}} _{i}}({\bf{x}})$, $i = 1,2,...,M$, via channel environment measurements, global positioning systems and so on. Under this setup, the secrecy outage probability, denoted as ${{\mathop{\rm P}\nolimits} _{{\rm{out}}}}$, is usually exploited to characterize the security metric. Specifically, ${{\mathop{\rm P}\nolimits} _{{\rm{out}}}}$ is defined as the probability that the targeted PLS coding rate of Alice's encoder, i.e., $R_s$, is larger than the secrecy rate $C_s$ \cite{RIS_outage_probability}. Hence, based on (8), ${{\mathop{\rm P}\nolimits} _{{\rm{out}}}}$ is derived as
\begin{equation}
\begin{split}{}
&{{\rm{P}}_{{\rm{out}}}} = \Pr \left( {{C_s} \le {R_s}} \right) = \Pr \left( {{C_e} \ge {C_m} - {R_s}} \right)\\
=&\Pr \left( {\sum\nolimits_{i = 1}^M {{{\left| {{{\bf{h}}_i}({\bf{x}}){\bf{w}}} \right|}^2}}  \ge \frac{{{\sigma ^2}}}{P_a}\left( {\frac{{{P_a}{{\left| {{{\bf{h}}_0}({\bf{x}}){\bf{w}}} \right|}^2}}}{{{\sigma ^2}{2^{{R_s}}}}} + \frac{1}{{{2^{{R_s}}}}} - 1} \right)} \right).
%
%
% =& \Pr \left( \begin{array}{l}
%\sum\nolimits_{i = 1}^M {{{\left| {{{\bf{h}}_i}({\bf{x}}){\bf{w}}} \right|}^2}}  \ge \frac{{{\sigma ^2}}}{P}\\
% \times \left( {\frac{{P{{\left| {{{\bf{h}}_0}({\bf{x}}){\bf{w}}} \right|}^2}}}{{{\sigma ^2}{2^{{R_s}}}}} + \frac{1}{{{2^{{R_s}}}}} - 1} \right)
%\end{array} \right).
\end{split}
\end{equation}

\begin{figure*}[b!]
  \vspace{-0pt}
   \hrulefill
\setcounter{mytempeqncnt}{\value{equation}}
\setcounter{equation}{18}
\begin{equation}
\begin{split}{}
&{\rm{P}}_{{\rm{out}}}= 1 - {F_{\sum\nolimits_{i = 1}^M {{{\left| {{{\bf{h}}_i}({\bf{x}}){\bf{w}}} \right|}^2}} }}\left( {\frac{{{\sigma ^2}}}{P_a}\left( {\frac{{{P_a}{{\left| {{{\bf{h}}_0}({\bf{x}}){\bf{w}}} \right|}^2}}}{{{\sigma ^2}{2^{{R_s}}}}} + \frac{1}{{{2^{{R_s}}}}} - 1} \right)} \right)\\
 =& 1 - \gamma \left( {\underbrace {\frac{{{{\left( {\sum\nolimits_{i = 1}^M {\frac{{{\beta _i}}}{{{K_i} + 1}}\left( {{K_i}{{\left| {{{\overline {\bf{h}} }_i}({\bf{x}}){\bf{w}}} \right|}^2} + 1} \right)} } \right)}^2}}}{{\sum\nolimits_{i = 1}^M {{{\left( {\frac{{{\beta _i}}}{{{K_i} + 1}}} \right)}^2}\left( {2{K_i}{{\left| {{{\overline {\bf{h}} }_i}({\bf{x}}){\bf{w}}} \right|}^2} + 1} \right)} }}}_{{f_1}({\bf{w}},{\bf{x}})},\underbrace {\frac{{\sum\nolimits_{i = 1}^M {\frac{{{\beta _i}}}{{{K_i} + 1}}\left( {{K_i}{{\left| {{{\overline {\bf{h}} }_i}({\bf{x}}){\bf{w}}} \right|}^2} + 1} \right)} }}{{\sum\nolimits_{i = 1}^M {{{\left( {\frac{{{\beta _i}}}{{{K_i} + 1}}} \right)}^2}\left( {2{K_i}{{\left| {{{\overline {\bf{h}} }_i}({\bf{x}}){\bf{w}}} \right|}^2} + 1} \right)} }}\frac{{{\sigma ^2}}}{{{P_a}}}\left( {\frac{{{P_a}{{\left| {{{\bf{h}}_0}({\bf{x}}){\bf{w}}} \right|}^2}}}{{{\sigma ^2}{2^{{R_s}}}}} + \frac{1}{{{2^{{R_s}}}}} - 1} \right)}_{{f_2}({\bf{w}},{\bf{x}})}} \right).
\end{split}
\end{equation}
\setcounter{equation}{\value{mytempeqncnt}}
%\vspace{-5pt}
\end{figure*}

\subsection{Problem Formulation}
In this paper, we aim to minimize the secrecy outage probability ${\rm{P}}_{\rm{out}}$, by jointly optimizing the beamforming vector ${\bf{w}}$ and antennas' positions ${\bf{x}}$ at Alice. The optimization problem can be formulated as
 \begin{align}
&({\rm{P1}}):{\rm{  }}\mathop {\min }\limits_{{{\bf{w}},{\bf{x}}}} \ {\rm{P}}_{\rm{out}} \tag{${\rm{10a}}$}\\
{\rm{              }}&\ {\rm{s.t.}} \quad \left\| {\bf{w}} \right\| = 1,\tag{${\rm{10b}}$}\\
&\quad \quad \ \ \ {\bf{x}} \in {\cal C},\tag{${\rm{10c}}$}
%& \ \ \ \quad {1_{\mathbb{C}}} = 1,\ {\rm{if}} \ \\
%&0 < {Q_E} \le \max \left( {{Q_{th}}/\left( {\frac{1}{{{\lambda _{EE}}}} + \frac{{{\beta %^r}N}}{{{\lambda _{RE}}{\lambda _{ER}}}}} \right),{Q_{E,\max }}} \right)\tag{${\rm{11d}}$}.
 \end{align}
where ${\cal C}$ in (10c) denotes the feasible moving region for $N$ antennas at Alice. Similar to \cite{Guojie_Fluid}, in light of the following two aspects: i) there should exist a minimum distance, denoted as $d_{{\rm{min}}}$, between any two adjacent MAs to avoid the coupling effect, i.e., $\left| {{x_i} - {x_j}} \right| \ge {d_{\min }}$, $\forall i \ne j$; ii) the movement span should be the same for each antenna, we can set ${\cal C}$ as ${\cal C} \buildrel \Delta \over = \left\{ {{x_i} \in [{F_i},{G_i}]} \right\}_{i = 1}^N$, where
\begin{equation}
\setcounter{equation}{11}
\begin{split}{}
{F_i} =& \frac{{L - (N - 1){d_{\min }}}}{N}(i - 1) + (i - 1){d_{\min }},\\
{G_i} =& \frac{{L - (N - 1){d_{\min }}}}{N}i + (i - 1){d_{\min }},
\end{split}
\end{equation}
 based on which we have $0 = {F_1} < {G_1} < {F_2} < {G_2} < ... < {F_N} < {G_N} = L$ and ${G_i} - {F_i} = \frac{{L - (N - 1){d_{\min }}}}{N}$, $\forall i = 1,...,N$.

\textbf{Remark 1:} Given ${\bf{w}}$ and ${\bf{x}}$: i) the exact value of ${\left| {{{\bf{h}}_0}({\bf{x}}){\bf{w}}} \right|^2}$ is known and $R_s$ is a constant; ii) however, $\sum\nolimits_{i = 1}^M {{{\left| {{{\bf{h}}_i}({\bf{x}}){\bf{w}}} \right|}^2}} $ is totally random because Alice does not know the instantaneous values of ${\widetilde {\bf{h}}_i}$, $i = 1,2,...,M$. Therefore, to solve (P1), next we need to first derive the cumulative distribution function (CDF) of $\sum\nolimits_{i = 1}^M {{{\left| {{{\bf{h}}_i}({\bf{x}}){\bf{w}}} \right|}^2}} $ to pave the way for obtaining the expression of ${\rm{P}}_{{\rm{out}}}$.

\section{Analysis About the Expression of ${\rm{P}}_{{\rm{out}}}$}
\subsection{The CDF of $\sum\nolimits_{i = 1}^M {{{\left| {{{\bf{h}}_i}({\bf{x}}){\bf{w}}} \right|}^2}} $}
Given ${\bf{w}}$ and ${\bf{x}}$, based on (1) we can expand $\sum\nolimits_{i = 1}^M {{{\left| {{{\bf{h}}_i}({\bf{x}}){\bf{w}}} \right|}^2}} $ as
\begin{equation}
\begin{split}{}
&\sum\nolimits_{i = 1}^M {{{\left| {{{\bf{h}}_i}({\bf{x}}){\bf{w}}} \right|}^2}} \\
 =& \sum\nolimits_{i = 1}^M {{{\left| {\underbrace {\sqrt {\frac{{{K_i}{\beta _i}}}{{{K_i} + 1}}} {{\overline {\bf{h}} }_i}({\bf{x}}){\bf{w}}}_{{\rm{deterministic, \ denoted \ as }}{{\ \rm{X}}_i}} + \underbrace {\sqrt {\frac{{{\beta _i}}}{{{K_i} + 1}}} {{\widetilde {\bf{h}}}_i}{\bf{w}}}_{{\rm{random, \ denoted \ as }}{{\ \rm{Y}}_i}}} \right|}^2}} .
\end{split}
\end{equation}
Observing (12), each ${{{\bf{h}}_i}({\bf{x}}){\bf{w}}}$, $i = 1,2,...,M$, consists of a deterministic component ${\rm{X}}_i$ and a circularly-symmetric complex Gaussian random component ${\rm{Y}}_i$ (since ${\bf{w}}$ is independent of ${{{\widetilde {\bf{h}}}_i}}$). Hence, it is concluded that ${\left| {{{\bf{h}}_i}({\bf{x}}){\bf{w}}} \right|}$ follows a Rician distribution, where the total power (denoted as ${\widetilde Z_i}$) and the Rician K-factor (denoted as ${\widetilde K_i}$) are
\begin{equation}
\begin{split}{}
{\widetilde Z_i} =& {\left| {{{\rm{X}}_i}} \right|^2} + {\mathbb{E}}\left[ {{{\left| {{{\rm{Y}}_i}} \right|}^2}} \right]\mathop  = \limits^{(a)} \frac{{{K_i}{\beta _i}{{\left| {{{\overline {\bf{h}} }_i}({\bf{x}}){\bf{w}}} \right|}^2}}}{{{K_i} + 1}} + \frac{{{\beta _i}}}{{{K_i} + 1}},\\
{\widetilde K_i} =& \frac{{{{\left| {{{\rm{X}}_i}} \right|}^2}}}{{{\mathbb{E}}\left[ {{{\left| {{{\rm{Y}}_i}} \right|}^2}} \right]}} = {K_i}{\left| {{{\overline {\bf{h}} }_i}({\bf{x}}){\bf{w}}} \right|^2},
\end{split}
\end{equation}
and the equality $\mathop  = \limits^{(a)} $ is established because ${\mathbb{E}}\left[ {{{\left| {{{\rm{Y}}_i}} \right|}^2}} \right] = \frac{{{\beta _i}}}{{{K_i} + 1}}{\mathbb{E}}\left[ {{{\left| {{{\widetilde {\bf{h}}}_i}{\bf{w}}} \right|}^2}} \right] = \frac{{{\beta _i}}}{{{K_i} + 1}}$.

It is well-known that the CDF of the Rician random variable involves the standard Marcum-Q function, which is not favorable for subsequent analysis. To overcome this challenge, same as \cite{Yanshihao_TWC}, we can conveniently regard the Rician fading as a special case of Nakagami fading, based on which the probability density function (PDF) of ${{{\left| {{{\bf{h}}_i}({\bf{x}}){\bf{w}}} \right|}^2}}$ can be tightly approximated as
\begin{equation}
\begin{split}{}
{f_{{{\left| {{{\bf{h}}_{i}}({\bf{x}}){\bf{w}}} \right|}^2}}}(x) =&  {\left( {\frac{{\widetilde {{m_i}}}}{{\widetilde {{Z_i}}}}} \right)^{\widetilde {{m_i}}}}\frac{{{x^{\widetilde {{m_i}} - 1}}}}{{\Gamma (\widetilde {{m_i}})}}{e^{ - \frac{{\widetilde {{m_i}}}}{{\widetilde {{Z_i}}}}x}},%\\
%{F_{{{\left| {{{\bf{h}}_{i}}({\bf{x}}){\bf{w}}} \right|}^2}}}(x) =& \int_0^x {{f_{{{\left| {{{\bf{h}}_{a{e_i}}}({\bf{x}}){\bf{w}}} \right|}^2}}}(t)dt}  = \gamma \left( {\widetilde {{m_i}},\frac{{\widetilde {{m_i}}}}{{\widetilde {{Z_i}}}}x} \right),
\end{split}
\end{equation}
where ${\widetilde {{m_i}}}$ is the Nakagami fading parameter denoted by $\widetilde {{m_i}} = {({\widetilde K_i} + 1)^2}/(2{\widetilde K_i} + 1)$. Since the PDF of ${{{\left| {{{\bf{h}}_i}({\bf{x}}){\bf{w}}} \right|}^2}}$ is known, its expectation and variance are easily derived as ${\mathbb{E}}\left[ {{{\left| {{{\bf{h}}_i}({\bf{x}}){\bf{w}}} \right|}^2}} \right] = \widetilde {{Z_i}}$ and ${\mathbb{V}}\left[ {{{\left| {{{\bf{h}}_i}({\bf{x}}){\bf{w}}} \right|}^2}} \right] = \frac{{{{\widetilde {{Z_i}}}^2}}}{{\widetilde {{m_i}}}}$, respectively.

Although the tight and simple-structure PDF of each ${{{\left| {{{\bf{h}}_{i}}({\bf{x}}){\bf{w}}} \right|}^2}}$ is derived successfully, the PDF of the sum of each ${{{\left| {{{\bf{h}}_{i}}({\bf{x}}){\bf{w}}} \right|}^2}}$, $i = 1,2,...,M$, is still hard to be obtained. To make process, we now approximate the PDF of ${\sum\nolimits_{i = 1}^M {{{\left| {{{\bf{h}}_i}({\bf{x}}){\bf{w}}} \right|}^2}} }$ as a Gamma distribution by exploiting the Laguerre series approximation \cite{application}, i.e.,
\begin{equation}
{f_{\sum\nolimits_{i = 1}^M {{{\left| {{{\bf{h}}_i}({\bf{x}}){\bf{w}}} \right|}^2}} }}(x) \approx \frac{1}{{\Gamma (\mu ){\vartheta ^\mu }}}{x^{\mu  - 1}}{e^{ - \frac{x}{\vartheta }}},
\end{equation}
where $\mu $ and $\vartheta $ are the shape parameter and scale parameter, which are defined as
\begin{equation}
\mu  \buildrel \Delta \over = \frac{{{{\left( {{\mathbb{E}}\left[ {\sum\nolimits_{i = 1}^M {{{\left| {{{\bf{h}}_i}({\bf{x}}){\bf{w}}} \right|}^2}} } \right]} \right)}^2}}}{{{\mathbb{V}}\left[ {\sum\nolimits_{i = 1}^M {{{\left| {{{\bf{h}}_i}({\bf{x}}){\bf{w}}} \right|}^2}} } \right]}},\vartheta  \buildrel \Delta \over = \frac{{{\mathbb{V}}\left[ {\sum\nolimits_{i = 1}^M {{{\left| {{{\bf{h}}_i}({\bf{x}}){\bf{w}}} \right|}^2}} } \right]}}{{{\mathbb{E}}\left[ {\sum\nolimits_{i = 1}^M {{{\left| {{{\bf{h}}_i}({\bf{x}}){\bf{w}}} \right|}^2}} } \right]}},
\end{equation}
 and ${{\mathbb{E}}\left[ {\sum\nolimits_{i = 1}^M {{{\left| {{{\bf{h}}_i}({\bf{x}}){\bf{w}}} \right|}^2}} } \right]}$ and ${{\mathbb{V}}\left[ {\sum\nolimits_{i = 1}^M {{{\left| {{{\bf{h}}_i}({\bf{x}}){\bf{w}}} \right|}^2}} } \right]}$ are computed as
\begin{equation}
\begin{split}{}
&{\mathbb{E}}\left[ {\sum\nolimits_{i = 1}^M {{{\left| {{{\bf{h}}_i}({\bf{x}}){\bf{w}}} \right|}^2}} } \right] = \sum\nolimits_{i = 1}^M {{\mathbb{E}}\left[ {{{\left| {{{\bf{h}}_i}({\bf{x}}){\bf{w}}} \right|}^2}} \right]}  = \sum\nolimits_{i = 1}^M {\widetilde {{Z_i}}}, \\
&{\mathbb{V}}\left[ {\sum\nolimits_{i = 1}^M {{{\left| {{{\bf{h}}_i}({\bf{x}}){\bf{w}}} \right|}^2}} } \right] = \sum\nolimits_{i = 1}^M {{\mathbb{V}}\left[ {{{\left| {{{\bf{h}}_i}({\bf{x}}){\bf{w}}} \right|}^2}} \right]}  = \sum\nolimits_{i = 1}^M {\frac{{{{\widetilde {{Z_i}}}^2}}}{{\widetilde {{m_i}}}}}.
\end{split}
\end{equation}

\begin{figure}
\centering
\includegraphics[width=7.4cm]{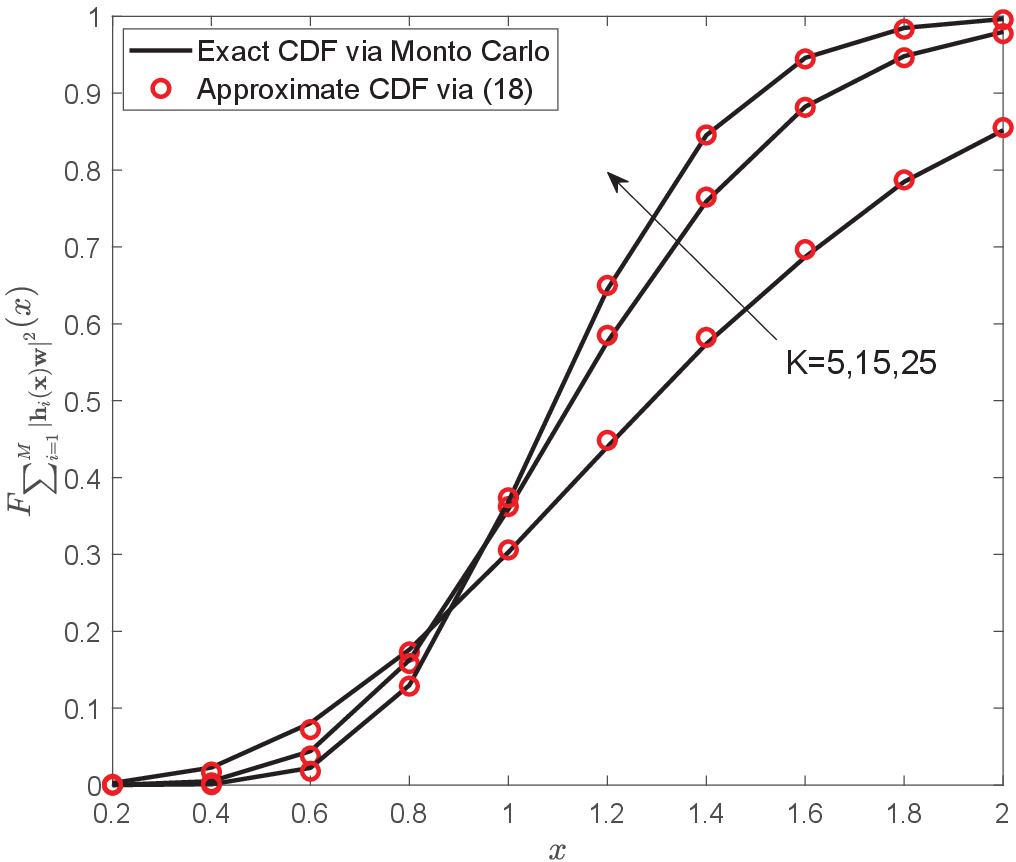}
\captionsetup{font=small}
\caption{Comparison of the exact CDF of ${\sum\nolimits_{i = 1}^M {{{\left| {{{\bf{h}}_i}({\bf{x}}){\bf{w}}} \right|}^2}} }$ via Monto Carlo and the derived approximation via (18), where $N = 5$, $M = 3$, ${\bf{x}} = {[0,0.5\lambda ,\lambda ,1.5\lambda ,2\lambda ]^T}$, ${\bf{w}} = {{\bf{w}}_1}/\left\| {{{\bf{w}}_1}} \right\|$ with ${{\bf{w}}_1} = {\left[ {1 + j,2 + j3,2 + j,3 - j,4 + j5} \right]^T}$, $\left\{ {{\beta _i}} \right\}_{i = 1}^M = 1$, ${\theta _1} = \pi /6$, ${\theta _2} = \pi /4$, ${\theta _3} = \pi /10$ and $\left\{ {{K_i}} \right\}_{i = 1}^M = K$.} \label{fig:Fig1}
%\vspace{-15pt}
\end{figure}

Substituting (17) into (16) and then substituting (16) into (15), the CDF of ${\sum\nolimits_{i = 1}^M {{{\left| {{{\bf{h}}_i}({\bf{x}}){\bf{w}}} \right|}^2}} }$ can be derived as
\begin{equation}
\begin{split}{}
{F_{\sum\nolimits_{i = 1}^M {{{\left| {{{\bf{h}}_i}({\bf{x}}){\bf{w}}} \right|}^2}} }}(x) =& \int_0^x {{f_{\sum\nolimits_{i = 1}^M {{{\left| {{{\bf{h}}_i}({\bf{x}}){\bf{w}}} \right|}^2}} }}(t)dt} \\
 =& \gamma \left( {\frac{{{{\left( {\sum\nolimits_{i = 1}^M {\widetilde {{Z_i}}} } \right)}^2}}}{{\sum\nolimits_{i = 1}^M {\frac{{{{\widetilde {{Z_i}}}^2}}}{{\widetilde {{m_i}}}}} }},\frac{{\sum\nolimits_{i = 1}^M {\widetilde {{Z_i}}} }}{{\sum\nolimits_{i = 1}^M {\frac{{{{\widetilde {{Z_i}}}^2}}}{{\widetilde {{m_i}}}}} }}x} \right),
\end{split}
\end{equation}
where $\gamma (a,x) = \frac{1}{{\Gamma (a)}}\int_0^x {{t^{a - 1}}{e^{ - t}}dt} $ is the lower incomplete gamma function.

As a supplement, we further compare the obtained ${F_{\sum\nolimits_{i = 1}^M {{{\left| {{{\bf{h}}_i}({\bf{x}}){\bf{w}}} \right|}^2}} }}(x)$ in (18) and the exact CDF (via Monto Carlo) as shown in Fig. 2, from which it is clear that our derived approximation is very tight and thus can be effectively exploited for subsequent optimization analysis.

\subsection{The Expression of ${\rm{P}}_{{\rm{out}}}$}
Based on (9) and the CDF of ${\sum\nolimits_{i = 1}^M {{{\left| {{{\bf{h}}_i}({\bf{x}}){\bf{w}}} \right|}^2}} }$ derived in (18), and by further expanding each ${\widetilde {{Z_i}}}$ and ${\widetilde {{m_i}}}$, $i = 1,2,...,M$, we can finally derive ${\rm{P}}_{{\rm{out}}}$ as in (19).

\textbf{Insight 1:} The secrecy outage probability ${\rm{P}}_{{\rm{out}}}$ is monotonically decreasing w.r.t. the transmit power of Alice and then converges to a constant when the power budget $P_a$ tends to infinity.

\begin{proof}
Observing (19), it is clear that ${{f_1}({\bf{w}},{\bf{x}})}$ is not related to $P_a$, while ${{f_2}({\bf{w}},{\bf{x}})}$ can be expanded as
\begin{equation}
\setcounter{equation}{20}
\begin{split}{}
&{f_2}({\bf{w}},{\bf{x}}) = \underbrace {\frac{{\sum\nolimits_{i = 1}^M {\frac{{{\beta _i}}}{{{K_i} + 1}}\left( {{K_i}{{\left| {{{\overline {\bf{h}} }_i}({\bf{x}}){\bf{w}}} \right|}^2} + 1} \right)} }}{{\sum\nolimits_{i = 1}^M {{{\left( {\frac{{{\beta _i}}}{{{K_i} + 1}}} \right)}^2}\left( {2{K_i}{{\left| {{{\overline {\bf{h}} }_i}({\bf{x}}){\bf{w}}} \right|}^2} + 1} \right)} }}\frac{{{{\left| {{{\bf{h}}_0}({\bf{x}}){\bf{w}}} \right|}^2}}}{{{2^{{R_s}}}}}}_{{\rm{constant}}}\\
 &+ \underbrace {\frac{{\sum\nolimits_{i = 1}^M {\frac{{{\beta _i}}}{{{K_i} + 1}}\left( {{K_i}{{\left| {{{\overline {\bf{h}} }_i}({\bf{x}}){\bf{w}}} \right|}^2} + 1} \right)} }}{{\sum\nolimits_{i = 1}^M {{{\left( {\frac{{{\beta _i}}}{{{K_i} + 1}}} \right)}^2}\left( {2{K_i}{{\left| {{{\overline {\bf{h}} }_i}({\bf{x}}){\bf{w}}} \right|}^2} + 1} \right)} }}\frac{{{\sigma ^2}}}{{{P_a}}}\left( {\frac{1}{{{2^{{R_s}}}}} - 1} \right)}_{{\rm{related}} \ {\rm{to}} \ {{P}_{{a}}}},
\end{split}
\end{equation}
from (20) it is easy to determine that ${{f_2}({\bf{w}},{\bf{x}})}$ is monotonically increasing w.r.t. $P_a$ (since $\frac{1}{{{2^{{R_s}}}}} - 1 < 0$). Further, note that $\gamma \left( {{f_1}({\bf{w}},{\bf{x}}),{f_2}({\bf{w}},{\bf{x}})} \right)$ is monotonically increasing w.r.t. ${{f_2}({\bf{w}},{\bf{x}})}$. Hence, it is always favorable to increase the transmit power of Alice to increase $\gamma \left( {{f_1}({\bf{w}},{\bf{x}}),{f_2}({\bf{w}},{\bf{x}})} \right)$ so as to decrease ${\rm{P}}_{{\rm{out}}} = 1 - \gamma \left( {{f_1}({\bf{w}},{\bf{x}}),{f_2}({\bf{w}},{\bf{x}})} \right)$. This is why we set the transmit power of Alice to its power budget in Section II.

 In addition, when ${P_a} \to \infty $, it has $\frac{{{\sigma ^2}}}{{{P_a}}} \to 0$, leading to
 \begin{equation}
\begin{split}{}
 {f_2}({\bf{w}},{\bf{x}}) \to \frac{{\sum\nolimits_{i = 1}^M {\frac{{{\beta _i}}}{{{K_i} + 1}}\left( {{K_i}{{\left| {{{\overline {\bf{h}} }_i}({\bf{x}}){\bf{w}}} \right|}^2} + 1} \right)} }}{{\sum\nolimits_{i = 1}^M {{{\left( {\frac{{{\beta _i}}}{{{K_i} + 1}}} \right)}^2}\left( {2{K_i}{{\left| {{{\overline {\bf{h}} }_i}({\bf{x}}){\bf{w}}} \right|}^2} + 1} \right)} }}\frac{{{{\left| {{{\bf{h}}_0}({\bf{x}}){\bf{w}}} \right|}^2}}}{{{2^{{R_s}}}}},
 \end{split}
\end{equation}
which is not related to $P_a$ anymore. Thus, when ${P_a} \to \infty $, ${\rm{P}}_{{\rm{out}}}$ will converge to a constant. This completes the proof.
\end{proof}

\textbf{Insight 2:} In the special case of $\left\{ {{K_i}} \right\}_{i = 1}^M = 0$, the wiretap channel suffers from Rayleigh fading and is totally random to Alice. Then, since there is no any useful information about the wiretap channel that can be exploited, to minimize ${\rm{P}}_{{\rm{out}}}$, Alice should just maximize the power gain of the main channel, i.e., maximize the term ${{{\left| {{{\bf{h}}_0}({\bf{x}}){\bf{w}}} \right|}^2}}$.
\begin{proof}
By substituting $\left\{ {{K_i}} \right\}_{i = 1}^M = 0$ into (19), ${\rm{P}}_{{\rm{out}}}$ can be simplified as ${\rm{P}}_{{\rm{out}}} = 1 - \gamma \left( {\frac{{{{\left( {\sum\nolimits_{i = 1}^M {{\beta _i}} } \right)}^2}}}{{\sum\nolimits_{i = 1}^M {\beta _i^2} }},\frac{{\sum\nolimits_{i = 1}^M {{\beta _i}} }}{{\sum\nolimits_{i = 1}^M {\beta _i^2} }}\frac{{{\sigma ^2}}}{{{P_a}}}\left( {\frac{{{P_a}{{\left| {{{\bf{h}}_0}({\bf{x}}){\bf{w}}} \right|}^2}}}{{{\sigma ^2}{2^{{R_s}}}}} + \frac{1}{{{2^{{R_s}}}}} - 1} \right)} \right)$, based on which the conclusion in Insight 2 can be easily obtained. This completes the proof.
\end{proof}

 Based on the above analysis, in the case of $\left\{ {{K_i}} \right\}_{i = 1}^M = 0$, we should just consider the following optimization problem to minimize ${\rm{P}}_{{\rm{out}}}$:
\begin{align}
&({\rm{P2}}):{\rm{  }}\mathop {\max }\limits_{{{\bf{w}},{\bf{x}}}} \ {{{\left| {{{\bf{h}}_0}({\bf{x}}){\bf{w}}} \right|}^2}} \tag{${\rm{22a}}$}\\
{\rm{              }}&\ {\rm{s.t.}} \quad (10{\rm{b}}), (10{\rm{c}}).\tag{${\rm{22b}}$}
%& \ \ \ \quad {1_{\mathbb{C}}} = 1,\ {\rm{if}} \ \\
%&0 < {Q_E} \le \max \left( {{Q_{th}}/\left( {\frac{1}{{{\lambda _{EE}}}} + \frac{{{\beta %^r}N}}{{{\lambda _{RE}}{\lambda _{ER}}}}} \right),{Q_{E,\max }}} \right)\tag{${\rm{11d}}$}.
 \end{align}
Problem (P2) can be easily solved by noting that given any ${\bf{x}} \in {{\cal C}}$, the optimal ${\bf{w}}$ is in the form of maximum-ratio transmission (MRT), i.e., ${{\bf{w}}^*} = {\bf{h}}_0^H({\bf{x}})/\left\| {{\bf{h}}_0^H({\bf{x}})} \right\|$, where ${\left(  \cdot  \right)^H}$ is the conjugate transpose operation. Based on this, we have ${\left| {{{\bf{h}}_0}({\bf{x}}){{\bf{w}}^*}} \right|^2} = {\left| {\frac{{{{\bf{h}}_0}({\bf{x}}){\bf{h}}_0^H({\bf{x}})}}{{\left\| {{\bf{h}}_0^H({\bf{x}})} \right\|}}} \right|^2} = {\beta _0}N$ given any ${\bf{x}} \in {{\cal C}}$. That is to say, in the case of $\left\{ {{K_i}} \right\}_{i = 1}^M = 0$, positions of antennas at Alice can be arbitrary in the specified region and Alice just employs the MRT-based beamforming for maximizing the receiving SNR of Bob.\footnotemark \footnotetext{Although the MRT-based beamforming is optimal when $\left\{ {{K_i}} \right\}_{i = 1}^M = 0$, it will achieve undesirable performance under the general case of $\left\{ {{K_i}} \right\}_{i = 1}^M > 0$ as pointed out in numerous literature and also shown in subsequent simulations. Therefore, we will not conduct further theoretical analysis on this scheme.}

\textbf{Insight 3:} In the special case of $\left\{ {{K_i}} \right\}_{i = 1}^M \to \infty $, i.e., the wiretap channel only consists of the LoS component and is totally deterministic to Alice, absolute secrecy can be guaranteed given $N \ge M + 1$.
\begin{proof}
When $\left\{ {{K_i}} \right\}_{i = 1}^M \to \infty $, ${{\bf{h}}_i}({\bf{x}})$ is simplified as ${\beta _i}{\overline {\bf{h}} _i}({\bf{x}})$, $i = 1,2,...,M$. Then, given $N \ge M + 1$, Alice can employ the simple ZF-based transmit beamforming (denoted as ${{{\bf{w}}_{{\rm{ZF}}}}}$) to completely eliminate the receiving power gain at each eavesdropper, i.e., ${\sum\nolimits_{i = 1}^M {\left| {{\beta _i}{{\overline {\bf{h}} }_i}({\bf{x}}){{\bf{w}}_{{\rm{ZF}}}}} \right|} ^2} = 0$. Then, based on (9) it is clear that ${P_{{\rm{out}}}} = \left( {0 \ge \frac{{{\sigma ^2}}}{{{P_a}}}\left( {\frac{{{P_a}{{\left| {{{\bf{h}}_0}({\bf{x}}){{\bf{w}}_{{\rm{ZF}}}}} \right|}^2}}}{{{\sigma ^2}{2^{{R_s}}}}} + \frac{1}{{{2^{{R_s}}}}} - 1} \right)} \right) = 0$. This completes the proof.
\end{proof}

\section{Novel Optimization Design to (P1)}
We now focus on how to efficiently solve (P1). Based on (19), minimizing ${\rm{P}}_{{\rm{out}}}$ is equivalent to maximizing $\gamma \left( {{f_1}({\bf{w}},{\bf{x}}),{f_2}({\bf{w}},{\bf{x}})} \right)$. Therefore, (P1) can be reformulated as
\begin{align}
&({\rm{P3}}):{\rm{  }}\mathop {\max }\limits_{{{\bf{w}},{\bf{x}}}} \ \gamma \left( {{f_1}({\bf{w}},{\bf{x}}),{f_2}({\bf{w}},{\bf{x}})} \right) \tag{${\rm{23a}}$}\\
{\rm{              }}&\ {\rm{s.t.}} \quad (10{\rm{b}}), (10{\rm{c}}).\tag{${\rm{23b}}$}
%& \ \ \ \quad {1_{\mathbb{C}}} = 1,\ {\rm{if}} \ \\
%&0 < {Q_E} \le \max \left( {{Q_{th}}/\left( {\frac{1}{{{\lambda _{EE}}}} + \frac{{{\beta %^r}N}}{{{\lambda _{RE}}{\lambda _{ER}}}}} \right),{Q_{E,\max }}} \right)\tag{${\rm{11d}}$}.
 \end{align}

In most literature, for the objective $\gamma \left( {a,b} \right)$, the optimization variables are just involved in the parameter $a$ or $b$. Then, maximizing $\gamma \left( {a,b} \right)$ can be conveniently transformed into minimizing $a$ or maximizing $b$, since $\gamma \left( {a,b} \right)$ is monotonically decreasing and increasing w.r.t. $a$ and $b$, respectively. With this way, the complex incomplete gamma function can be removed from the objective and then the optimization becomes much easier. However, observing the objective of (P3), the variables ${\bf{w}}$ and ${\bf{x}}$ are involved in both $a$ and $b$, which indicates that the conventional method is not suitable anymore and we have to seek new methods to solve (P3).

\begin{figure}
\centering
\includegraphics[width=7.2cm]{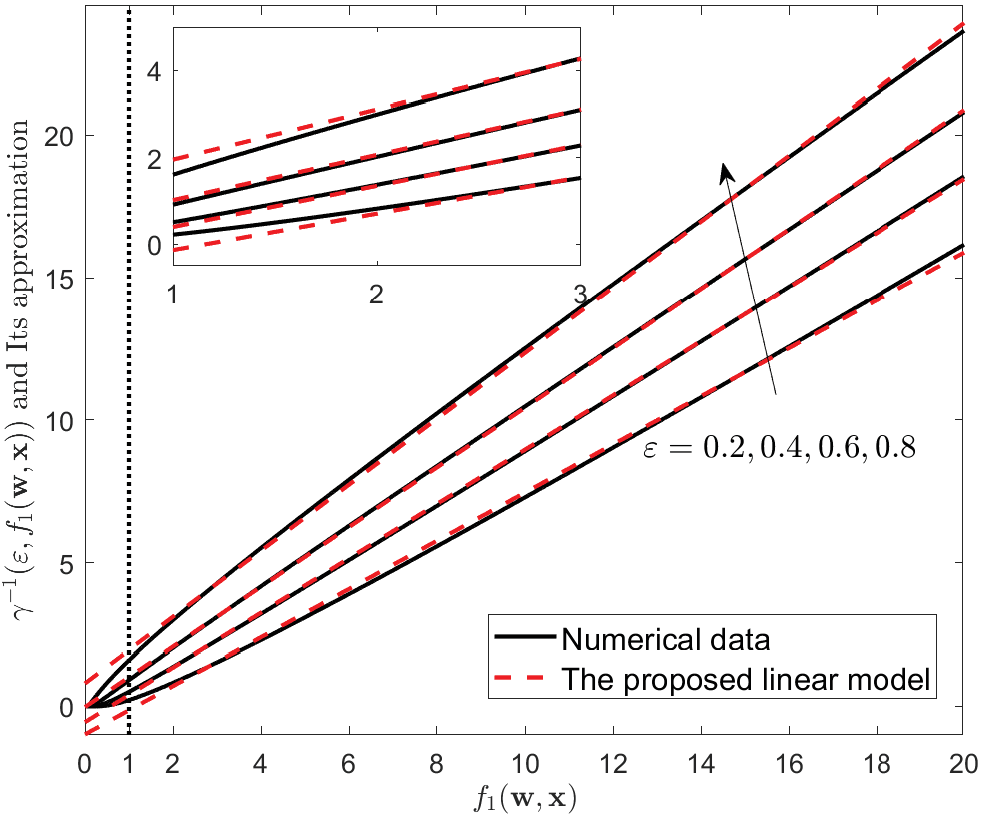}
\captionsetup{font=small}
\caption{${\gamma ^{ - 1}}(\varepsilon ,{f_1}({\bf{w}},{\bf{x}}))$ and the proposed linear approximation model w.r.t. ${f_1}({\bf{w}},{\bf{x}})$ under different $\varepsilon $.} \label{fig:Fig1}
%\vspace{-15pt}
\end{figure}

 \begin{figure}
\centering
\includegraphics[width=7.2cm]{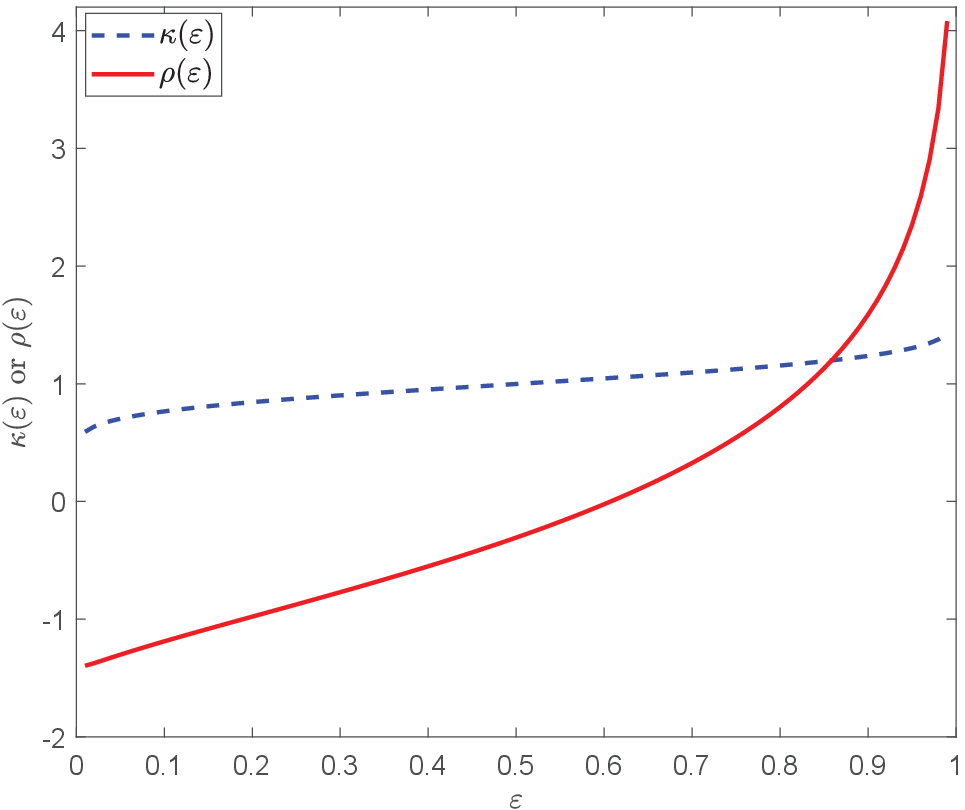}
\captionsetup{font=small}
\caption{The parameters $\kappa (\varepsilon )$ and $\rho (\varepsilon )$ w.r.t. $\varepsilon $.} \label{fig:Fig1}
%\vspace{-15pt}
\end{figure}

\subsection{A Novel Optimization Framework}
To proceed, we now introduce the slack variable $\varepsilon $ and equivalently express (P3) as
\begin{align}
&({\rm{P3.1}}):{\rm{  }}\mathop {\max }\limits_{{{\bf{w}},{\bf{x}}},\varepsilon } \ \varepsilon  \tag{${\rm{24a}}$}\\
{\rm{              }}&\ {\rm{s.t.}} \quad \ \gamma \left( {{f_1}({\bf{w}},{\bf{x}}),{f_2}({\bf{w}},{\bf{x}})} \right) \ge \varepsilon ,\tag{${\rm{24b}}$}\\
&\quad \quad \ \ \ (10{\rm{b}}), (10{\rm{c}}).\tag{${\rm{24c}}$}
%& \ \ \ \quad {1_{\mathbb{C}}} = 1,\ {\rm{if}} \ \\
%&0 < {Q_E} \le \max \left( {{Q_{th}}/\left( {\frac{1}{{{\lambda _{EE}}}} + \frac{{{\beta %^r}N}}{{{\lambda _{RE}}{\lambda _{ER}}}}} \right),{Q_{E,\max }}} \right)\tag{${\rm{11d}}$}.
 \end{align}

Note that the constraint (24b) is equivalent to
\begin{equation}
\setcounter{equation}{25}
\begin{split}{}
{f_2}({\bf{w}},{\bf{x}}) \ge {\gamma ^{ - 1}}\left( {\varepsilon ,{f_1}({\bf{w}},{\bf{x}})} \right),
\end{split}
\end{equation}
where ${\gamma ^{ - 1}}(\varepsilon ,{f_1}({\bf{w}},{\bf{x}}))$, with $\varepsilon  \in [0,1]$, is the inverse of the lower incomplete gamma function.

 \begin{figure*}
\centering
\includegraphics[width=7.2cm]{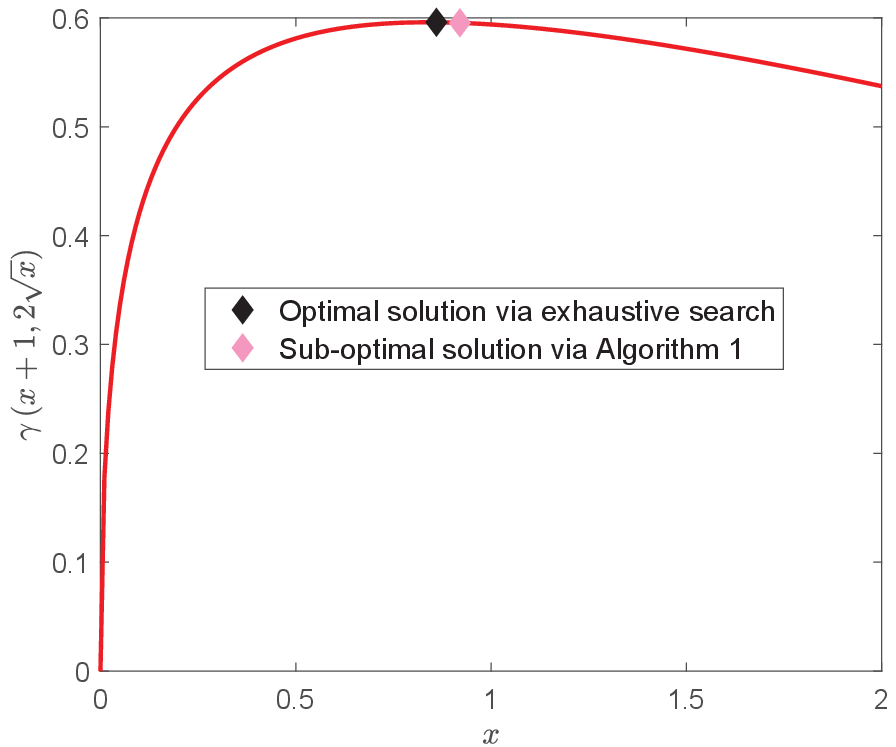}
\includegraphics[width=7.2cm]{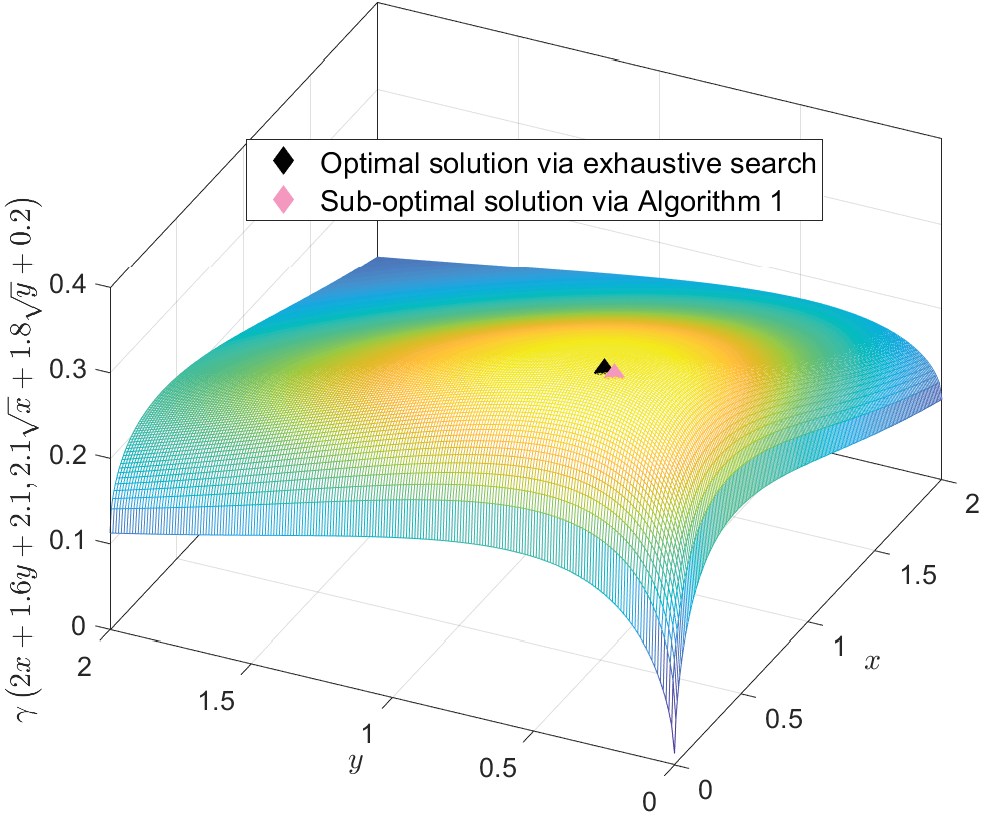}
\captionsetup{font=small}
\caption{The achievable performance of the proposed algorithm under two typical cases.} \label{fig:Fig1}
\vspace{-15pt}
\end{figure*}

Compared to the complex $\gamma \left( {{f_1}({\bf{w}},{\bf{x}}),{f_2}({\bf{w}},{\bf{x}})} \right)$, ${\gamma ^{ - 1}}\left( {\varepsilon ,{f_1}({\bf{w}},{\bf{x}})} \right)$ is only related to ${{f_1}({\bf{w}},{\bf{x}})}$ when $\varepsilon $ is given. This motivates us to plot the curves of ${\gamma ^{ - 1}}(\varepsilon ,{f_1}({\bf{w}},{\bf{x}}))$ w.r.t. ${f_1}({\bf{w}},{\bf{x}})$ under different $\varepsilon  \in [0,1]$ as shown in Fig. 3 (solid lines). From Fig. 3, we observe that ${\gamma ^{ - 1}}(\varepsilon ,{f_1}({\bf{w}},{\bf{x}}))$ almost increases linearly w.r.t. ${f_1}({\bf{w}},{\bf{x}})$ for any $\varepsilon $, especially when ${f_1}({\bf{w}},{\bf{x}})$ lies in a larger region. Based on this fact, the simple linear model (dashed lines in Fig. 3), i.e., $\kappa (\varepsilon ){f_1}({\bf{w}},{\bf{x}}) + \rho (\varepsilon )$, can be exploited to tightly approximate ${\gamma ^{ - 1}}(\varepsilon ,{f_1}({\bf{w}},{\bf{x}}))$, where $\kappa (\varepsilon )$ and $\rho (\varepsilon )$ are related to the parameter $\varepsilon $ and are determined based on the criterion of minimum mean square error. In particular,
the ``nlinfit'' function in Matlab can be used to find $\kappa (\varepsilon )$ and $\rho (\varepsilon )$. We further plot $\kappa (\varepsilon )$ and $\rho (\varepsilon )$ w.r.t. $\varepsilon$ as shown in Fig. 4, from which $\kappa (\varepsilon )$ and $\rho (\varepsilon )$ are both monotonically increasing w.r.t. $\varepsilon $. In addition, $\kappa (\varepsilon ) > 0$, $\forall \varepsilon  \in [0,1]$, while $\rho (\varepsilon )$ can be positive or negative when $\varepsilon  \in [0,1]$.

\textbf{Remark 2:} From Fig. 3, the accuracy of the proposed approximation mainly depends on the value of ${f_1}({\bf{w}},{\bf{x}})$. Specifically, when ${f_1}({\bf{w}},{\bf{x}})$ is larger than two, the proposed linear model almost approaches the numerical data; while there exists the slight deviation between ${\gamma ^{ - 1}}\left( {\varepsilon ,{f_1}({\bf{w}},{\bf{x}})} \right)$ and $\kappa (\varepsilon ){f_1}({\bf{w}},{\bf{x}}) + \rho (\varepsilon )$ when ${f_1}({\bf{w}},{\bf{x}})$ does not exceed one. Fortunately, observing (19), we can derive that
\begin{equation}
\begin{split}{}
{f_1}({\bf{w}},{\bf{x}}) >& \frac{{\sum\nolimits_{i = 1}^M {{{\left( {\frac{{{\beta _i}}}{{{K_i} + 1}}} \right)}^2}{{\left( {{K_i}{{\left| {{{\overline {\bf{h}} }_i}({\bf{x}}){\bf{w}}} \right|}^2} + 1} \right)}^2}} }}{{\sum\nolimits_{i = 1}^M {{{\left( {\frac{{{\beta _i}}}{{{K_i} + 1}}} \right)}^2}\left( {2{K_i}{{\left| {{{\overline {\bf{h}} }_i}({\bf{x}}){\bf{w}}} \right|}^2} + 1} \right)} }}\\
 >& \frac{{\sum\nolimits_{i = 1}^M {{{\left( {\frac{{{\beta _i}}}{{{K_i} + 1}}} \right)}^2}\left( {2{K_i}{{\left| {{{\overline {\bf{h}} }_i}({\bf{x}}){\bf{w}}} \right|}^2} + 1} \right)} }}{{\sum\nolimits_{i = 1}^M {{{\left( {\frac{{{\beta _i}}}{{{K_i} + 1}}} \right)}^2}\left( {2{K_i}{{\left| {{{\overline {\bf{h}} }_i}({\bf{x}}){\bf{w}}} \right|}^2} + 1} \right)} }} > 1.
\end{split}
\end{equation}
Hence, it is concluded that the proposed linear model can achieve a pretty good fitting result.

Armed with the above analysis, we now can provide an approximate problem for (P3.1), i.e.,
\begin{align}
&({\rm{P3.2}}):{\rm{  }}\mathop {\max }\limits_{{{\bf{w}},{\bf{x}}},\varepsilon } \ \varepsilon  \tag{${\rm{27a}}$}\\
{\rm{              }}&\ {\rm{s.t.}} \quad \ {f_2}({\bf{w}},{\bf{x}}) \ge \kappa (\varepsilon ){f_1}({\bf{w}},{\bf{x}}) + \rho (\varepsilon ),\tag{${\rm{27b}}$}\\
&\quad \quad \ \ \ (10{\rm{b}}), (10{\rm{c}}).\tag{${\rm{27c}}$}
%& \ \ \ \quad {1_{\mathbb{C}}} = 1,\ {\rm{if}} \ \\
%&0 < {Q_E} \le \max \left( {{Q_{th}}/\left( {\frac{1}{{{\lambda _{EE}}}} + \frac{{{\beta %^r}N}}{{{\lambda _{RE}}{\lambda _{ER}}}}} \right),{Q_{E,\max }}} \right)\tag{${\rm{11d}}$}.
 \end{align}

Since $\kappa (\varepsilon )$ and $\rho (\varepsilon )$ have no closed-form expressions, it is still hard to directly solve (P3.2). Because of this reason, the following steps should be followed to find the optimal solution of (P3.2):

1) Use bisection to search $\varepsilon $, with $\varepsilon  \in [0,1]$;

2) Given each $\varepsilon $, check the feasibility of the following problem:
\begin{align}
&({\rm{P3.3}}): {\rm{Find}} \ \ {\bf{w}},{\bf{x}} \tag{${\rm{28a}}$}\\
{\rm{              }}&\ {\rm{s.t.}} \quad \ (27{\rm{b}}),(27{\rm{c}}).\tag{${\rm{28b}}$}
%& \ \ \ \quad {1_{\mathbb{C}}} = 1,\ {\rm{if}} \ \\
%&0 < {Q_E} \le \max \left( {{Q_{th}}/\left( {\frac{1}{{{\lambda _{EE}}}} + \frac{{{\beta %^r}N}}{{{\lambda _{RE}}{\lambda _{ER}}}}} \right),{Q_{E,\max }}} \right)\tag{${\rm{11d}}$}.
 \end{align}

Specifically, if (P3.3) is feasible, the optimal solution of (P3.2) must be no smaller than the given $\varepsilon $. Thus, in the next search, the updated $\varepsilon $ should be larger than the given $\varepsilon $ of the last round. Otherwise, if (P3.3) is infeasible, the optimal solution of (P3.2) must not be larger than the given $\varepsilon $. Thus, in the next search, the updated $\varepsilon $ should be smaller than the given $\varepsilon $ of the last round. In addition, to judge that whether (P3.3) is feasible or not, it is equivalent to considering the following problem:
\begin{align}
&({\rm{P3.4}}):{\rm{  }}\mathop {\max }\limits_{{{\bf{w}},{\bf{x}}}} \ {f_2}({\bf{w}},{\bf{x}}) - \kappa (\varepsilon ){f_1}({\bf{w}},{\bf{x}}) - \rho (\varepsilon ) \tag{${\rm{29a}}$}\\
{\rm{              }}&\ {\rm{s.t.}} \quad \ (10{\rm{b}}), (10{\rm{c}}).\tag{${\rm{29b}}$}
%& \ \ \ \quad {1_{\mathbb{C}}} = 1,\ {\rm{if}} \ \\
%&0 < {Q_E} \le \max \left( {{Q_{th}}/\left( {\frac{1}{{{\lambda _{EE}}}} + \frac{{{\beta %^r}N}}{{{\lambda _{RE}}{\lambda _{ER}}}}} \right),{Q_{E,\max }}} \right)\tag{${\rm{11d}}$}.
 \end{align}
Given $\varepsilon $, if the optimal output of (P3.4) is not larger than zero, (P3.3) is infeasible. Otherwise, (P3.3) is feasible.

\begin{algorithm}
\caption{Steps for Solving (P3.2)}
  \begin{algorithmic}[1]

\State \textbf{{Input:}} $\kappa (\varepsilon )$ and $\rho (\varepsilon )$, $\varepsilon  = 0:\tau :1$, $\tau $ is the accuracy.

\State \textbf{Initialize:} ${\varepsilon _1} = 0.5$, ${\varepsilon _L} = 0$ and ${\varepsilon _R} = 1$.

\State \textbf{Repeat:}

%\State \quad For $k = 1:1:M/2 - 1$:

\State \quad Set $\varepsilon  = {\varepsilon _1}$ and solve (P3.3).

\State \quad If (P3.3) is infeasible, i.e., the optimal output of (P3.4) is smaller than zero, update ${\varepsilon _R} = {\varepsilon _1};{\varepsilon _1} = ({\varepsilon _L} + {\varepsilon _1})/2$;

\State \quad Otherwise, update ${\varepsilon _L} = {\varepsilon _1};{\varepsilon _1} = ({\varepsilon _1} + {\varepsilon _R})/2$.
%\State \quad End

\State \textbf{Until:} ${\varepsilon _1}$ converges to a stationary value.

\State \textbf{Output:} ${{\bf{w}}_{{\varepsilon _1}}}$, ${{\bf{x}}_{{\varepsilon _1}}}$ and the resulted secrecy outage probability ${{\rm{P}}_{{\rm{out}}}} = 1 - \gamma \left( {{f_1}({{\bf{w}}_{{\varepsilon _1}}},{{\bf{x}}_{{\varepsilon _1}}}),{f_2}({{\bf{w}}_{{\varepsilon _1}}},{{\bf{x}}_{{\varepsilon _1}}})} \right)$.
  \end{algorithmic}
\end{algorithm}

Armed with the above analysis, we can finally list the detailed steps for solving (P3.2) as shown in Algorithm 1. After obtaining the stationary ${\varepsilon _1}$ of (P3.2) and the corresponding ${\bf{w}}$ (denoted as ${{\bf{w}}_{{\varepsilon _1}}}$) and ${\bf{x}}$ (denoted as ${{\bf{x}}_{{\varepsilon _1}}}$), the actual secrecy outage probability can be computed as ${{\rm{P}}_{{\rm{out}}}} = 1 - \gamma \left( {{f_1}({{\bf{w}}_{{\varepsilon _1}}},{{\bf{x}}_{{\varepsilon _1}}}),{f_2}({{\bf{w}}_{{\varepsilon _1}}},{{\bf{x}}_{{\varepsilon _1}}})} \right)$.

\begin{figure*}[b!]
  \vspace{-0pt}
   \hrulefill
\setcounter{mytempeqncnt}{\value{equation}}
\setcounter{equation}{30}
\begin{equation}
\begin{split}{}
{f_3}({\bf{w}},{\bf{x}}) =& \sum\nolimits_{i = 1}^M {\frac{{{\beta _i}}}{{{K_i} + 1}}\left( {{K_i}{{\left| {{{\overline {\bf{h}} }_i}({\bf{x}}){\bf{w}}} \right|}^2} + 1} \right)\frac{{{\sigma ^2}}}{{{P_a}}}\left( {\frac{{{P_a}{{\left| {{{\bf{h}}_0}({\bf{x}}){\bf{w}}} \right|}^2}}}{{{\sigma ^2}{2^{{R_s}}}}} + \frac{1}{{{2^{{R_s}}}}} - 1} \right)} \\
 &- \kappa (\varepsilon ){\left( {\sum\nolimits_{i = 1}^M {\frac{{{\beta _i}}}{{{K_i} + 1}}\left( {{K_i}{{\left| {{{\overline {\bf{h}} }_i}({\bf{x}}){\bf{w}}} \right|}^2} + 1} \right)} } \right)^2} - \rho (\varepsilon )\sum\nolimits_{i = 1}^M {{{\left( {\frac{{{\beta _i}}}{{{K_i} + 1}}} \right)}^2}\left( {2{K_i}{{\left| {{{\overline {\bf{h}} }_i}({\bf{x}}){\bf{w}}} \right|}^2} + 1} \right)}.
\end{split}
\end{equation}
\setcounter{equation}{\value{mytempeqncnt}}
%\vspace{-5pt}
\end{figure*}

\textbf{Remark 3:} Before implementing Algorithm 1, we have to list ${\left\{ {\kappa (\varepsilon ),\rho (\varepsilon )} \right\}_{\varepsilon  = 0:\tau :1}}$ based on numerical simulations. However, we must emphasize that ${\left\{ {\kappa (\varepsilon ),\rho (\varepsilon )} \right\}_{\varepsilon  = 0:\tau :1}}$ are not related to $\left\{ {{f_1}({\bf{w}},{\bf{x}}),{f_2}({\bf{w}},{\bf{x}})} \right\}$. Thus, ${\left\{ {\kappa (\varepsilon ),\rho (\varepsilon )} \right\}_{\varepsilon  = 0:\tau :1}}$ are just computed only once and then saved forever for handling any different $\left\{ {{f_1}({\bf{w}},{\bf{x}}),{f_2}({\bf{w}},{\bf{x}})} \right\}$.

\textbf{Remark 4:} In Algorithm 1, the bisection method is used to search each $\varepsilon $, leading to the corresponding search complexity as $\log \left( {1/\tau } \right)$. Furthermore, compared to the original problem (P3.1), the complexity of solving (P3.4) would be much lower due to its clear structure. We will verify this fact via two examples as shown subsequently.

\textbf{Example 1:} Let us first focus on a simple problem, i.e., $\mathop {\max }\limits_x \gamma \left( {x + 1,2\sqrt x } \right), {\rm{s.t}}.,x \in [0,2]$. As $\gamma \left( {a,b} \right)$ is monotonically increasing w.r.t. $b$ but decreasing w.r.t. $a$, it is hard to find the closed-form and optimal $x$ that maximizes $\gamma \left( {x + 1,2\sqrt x } \right)$. Hence, the optimal solution should be obtained via the one-dimensional search. If the search accuracy is also $\tau $, the whole complexity is ${\cal{O}}\left( {1/{\tau}} \right)$. Nevertheless, using Algorithm 1, given $\varepsilon $, the objective of (P3.4) becomes $2\sqrt x  - \kappa (\varepsilon )(x + 1) - \rho (\varepsilon )$ which is clearly concave and thus can be efficiently solved using existing solvers, e.g., CVX. This indicates that the whole complexity of implementing Algorithm 1 is only ${\cal O}\left( {\log (1/\tau ) \times {1^{3.5}}} \right) = {\cal O}\left( {\log (1/\tau )} \right)$, where ${\cal O}\left( {{1^{3.5}}} \right)$ is the complexity of running (P3.4) given $\varepsilon $.

\textbf{Example 2:} Let us now focus on a more complex problem, i.e., $\mathop {\max }\limits_{x,y} \gamma \left( {2x + 1.6y + 2.1,2.1\sqrt x  + 1.8\sqrt y  + 0.2} \right)$, ${\rm{s}}.{\rm{t}}.,x,y \in [0,2]$. Due to the same reason mentioned above, the optimal solution should be obtained via the two-dimensional search with the complexity of ${\cal{O}}\left( {1/{\tau ^2}} \right)$. Nevertheless, using Algorithm 1, given $\varepsilon $, the objective of (P3.4) becomes $2.1\sqrt x  + 1.8\sqrt y  + 0.2 - \kappa (\varepsilon )(2x + 1.6y + 2.1) - \rho (\varepsilon )$ and is still concave. Hence, the whole complexity of implementing Algorithm 1 is only ${\cal {O}}\left( {\log (1/\tau ) \times {2^{3.5}}} \right)$, where ${\cal O}\left( {{2^{3.5}}} \right)$ is the complexity of running (P3.4) given $\varepsilon $.

Via Examples 1 and 2, it is evident that our proposed novel algorithm can reduce the implementation complexity significantly. For instance, when $\tau  = 0.01$, the complexities of the exhaustive search and Algorithm 1 for Example 1 are ${\cal {O}}\left( {100} \right)$ and ${\cal {O}}\left( {4.6} \right)$, and for Example 2 are ${\cal {O}}\left( {100^2} \right)$ and ${\cal {O}}\left( {50} \right)$, respectively. More importantly, our proposed algorithm can also achieve the pretty good performance. This is verified in Fig. 5, from which it is clear that the sub-optimal solution found by Algorithm 1 almost approaches the optimal solution found via the exhaustive search.

\subsection{The APGA for Solving (P3.4)}
Via the above analysis, the key step in Algorithm 1 is to solve (P3.4) given each $\varepsilon $. To proceed, we first express the objective of (P3.4) as
\begin{equation}
\setcounter{equation}{30}
\begin{split}
&{f_2}({\bf{w}},{\bf{x}}) - \kappa (\varepsilon ){f_1}({\bf{w}},{\bf{x}}) - \rho (\varepsilon )\\
 =& \frac{{{f_3}({\bf{w}},{\bf{x}})}}{{\sum\nolimits_{i = 1}^M {{{\left( {\frac{{{\beta _i}}}{{{K_i} + 1}}} \right)}^2}\left( {2{K_i}{{\left| {{{\overline {\bf{h}} }_i}({\bf{x}}){\bf{w}}} \right|}^2} + 1} \right)} }},
\end{split}
\end{equation}
where ${{f_3}({\bf{w}},{\bf{x}})}$ is shown in (31).

 Note that we just need to determine that whether the optimal output of (P3.4) is larger than zero or not. Keeping this in mind and since $\frac{1}{{\sum\nolimits_{i = 1}^M {{{\left( {\frac{{{\beta _i}}}{{{K_i} + 1}}} \right)}^2}\left( {2{K_i}{{\left| {{{\overline {\bf{h}} }_i}({\bf{x}}){\bf{w}}} \right|}^2} + 1} \right)} }} > 0$, (P3.4) can be transformed into the following problem with a clear structure:
 \begin{align}
&({\rm{P3.5}}):{\rm{  }}\mathop {\max }\limits_{{{\bf{w}},{\bf{x}}}} \ {{f_3}({\bf{w}},{\bf{x}})} \tag{${\rm{32a}}$}\\
{\rm{              }}&\ {\rm{s.t.}} \quad \ (10{\rm{b}}), (10{\rm{c}}).\tag{${\rm{32b}}$}
%& \ \ \ \quad {1_{\mathbb{C}}} = 1,\ {\rm{if}} \ \\
%&0 < {Q_E} \le \max \left( {{Q_{th}}/\left( {\frac{1}{{{\lambda _{EE}}}} + \frac{{{\beta %^r}N}}{{{\lambda _{RE}}{\lambda _{ER}}}}} \right),{Q_{E,\max }}} \right)\tag{${\rm{11d}}$}.
 \end{align}
 Then, we can conclude that, given $\varepsilon $, if the optimal output of (P3.5) is not larger than zero, (P3.3) is infeasible. Otherwise, (P3.3) is feasible.

Problem (P3.5) is highly non-convex because its objective is much complex, which is neither convex or concave w.r.t. ${\bf{w}}$ and ${\bf{x}}$, and thus may not be solved via standard convex optimization techniques. Responding to this difficulty, we in this paper try to exploit the APGA method \cite{RIS_MIMO_tsp} to solve (P3.5), since the APGA handles the unconstrained or constrained problems well and is not sensitive to concavity or convexity of the objective.

 \begin{algorithm}
\caption{The APGA for Solving (P3.5)}
  \begin{algorithmic}[1]

\State \textbf{{Input:}} ${{\bf{w}}^{(0)}}$, with $\left\| {{{\bf{w}}^{(0)}}} \right\| = 1$, ${{\bf{x}}^{(0)}} \in {\cal C}$, $t \leftarrow 0$, $\delta  > 0$ and $\chi  < 1$.

%\State \textbf{Initialize:} ${\varepsilon _1} = 0.5$, ${\varepsilon _L} = 0$ and ${\varepsilon _R} = 1$.

\State \textbf{Repeat:}

%\State \quad For $k = 1:1:M/2 - 1$:

\State \quad \textbf{Repeat:} \quad \ \ $\% \% \% \% \%$  \textit{The steps for updating} ${\bf{w}}$

\State \quad \ ${{\bf{w}}^{(t + 1)}} = {{\bf{w}}^{(t)}} + \delta {\nabla _{\bf{w}}}{f_3}({{\bf{w}}^{(t)}},{{\bf{x}}^{(t)}})$;

\State \quad \ ${{\bf{w}}^{(t + 1)}} = {{\bf{w}}^{(t + 1)}}/\left\| {{{\bf{w}}^{(t + 1)}}} \right\|$;

\State \quad \ {\textbf{if}} ${f_3}({{\bf{w}}^{(t + 1)}},{{\bf{x}}^{(t)}}) < {Q_\delta }({{\bf{w}}^{(t + 1)}},{{\bf{x}}^{(t)}};{{\bf{w}}^{(t)}})$, {\textbf{then}}

\State \quad \quad \ $\delta  \leftarrow \chi \delta $,

\State \quad \ \textbf{end}

\State \quad \textbf{Until:} ${f_3}({{\bf{w}}^{(t + 1)}},{{\bf{x}}^{(t)}}) \ge {Q_\delta }({{\bf{w}}^{(t + 1)}},{{\bf{x}}^{(t)}};{{\bf{w}}^{(t)}})$.
%\State \quad For $k = 1:1:M/2 - 1$:

\State \quad \textbf{Repeat:} \quad \quad $\% \% \% \% \%$  \textit{The steps for updating} ${\bf{x}}$

\State \quad \ ${{\bf{x}}^{(t + 1)}} = {{\bf{x}}^{(t)}} + \delta {\nabla _{\bf{x}}}{f_3}({{\bf{w}}^{(t + 1)}},{{\bf{x}}^{(t)}})$;

\State \quad \ ${{\bf{x}}^{(t + 1)}} = {{\cal B}}\left\{ {{{\bf{x}}^{(t + 1)}},{{\cal C}}} \right\}$;

\State \quad \ {\textbf{if}} ${f_3}({{\bf{w}}^{(t + 1)}},{{\bf{x}}^{(t + 1)}}) < {Q_\delta }({{\bf{w}}^{(t + 1)}},{{\bf{x}}^{(t + 1)}};{{\bf{x}}^{(t)}})$, {\textbf{then}}

\State \quad \quad \ $\delta  \leftarrow \chi \delta $,

\State \quad \ \textbf{end}

\State \quad \textbf{Until:} ${f_3}({{\bf{w}}^{(t + 1)}},{{\bf{x}}^{(t + 1)}}) \ge {Q_\delta }({{\bf{w}}^{(t + 1)}},{{\bf{x}}^{(t + 1)}};{{\bf{x}}^{(t)}})$.

\State \quad $t \leftarrow t + 1$;

\State \textbf{Until:} ${f_3}({{\bf{w}}^{(t)}},{{\bf{x}}^{(t)}})$ converges to a stationary point.
  \end{algorithmic}
\end{algorithm}

 \begin{figure*}[b!]
  \vspace{-0pt}
   \hrulefill
\setcounter{mytempeqncnt}{\value{equation}}
\setcounter{equation}{34}
\begin{equation}
\begin{split}{}
{\nabla _{\bf{w}}}{f_3}({\bf{w}},{\bf{x}}) =& \sum\nolimits_{i = 1}^M {\frac{{{\beta _i}}}{{{K_i} + 1}}{K_i}\overline {\bf{h}} _i^H({\bf{x}}){{\overline {\bf{h}} }_i}({\bf{x}}){\bf{w}}} \frac{{{\sigma ^2}}}{{{P_a}}}\left( {\frac{{{P_a}{{\left| {{{\bf{h}}_0}({\bf{x}}){\bf{w}}} \right|}^2}}}{{{\sigma ^2}{2^{{R_s}}}}} + \frac{1}{{{2^{{R_s}}}}} - 1} \right)\\
 +& \sum\nolimits_{i = 1}^M {\frac{{{\beta _i}}}{{{K_i} + 1}}\left( {{K_i}{{\left| {{{\overline {\bf{h}} }_i}({\bf{x}}){\bf{w}}} \right|}^2} + 1} \right)} \frac{{{\bf{h}}_0^H({\bf{x}}){{\bf{h}}_0}({\bf{x}}){\bf{w}}}}{{{2^{{R_s}}}}} - 2\kappa (\varepsilon )\sum\nolimits_{i = 1}^M {\frac{{{\beta _i}}}{{{K_i} + 1}}\left( {{K_i}{{\left| {{{\overline {\bf{h}} }_i}({\bf{x}}){\bf{w}}} \right|}^2} + 1} \right)}  \\
 \times & \sum\nolimits_{i = 1}^M {\frac{{{\beta _i}}}{{{K_i} + 1}}{K_i}\overline {\bf{h}} _i^H({\bf{x}}){{\overline {\bf{h}} }_i}({\bf{x}}){\bf{w}}}  - 2\rho (\varepsilon )\sum\nolimits_{i = 1}^M {{{\left( {\frac{{{\beta _i}}}{{{K_i} + 1}}} \right)}^2}{K_i}\overline {\bf{h}} _i^H({\bf{x}}){{\overline {\bf{h}} }_i}({\bf{x}}){\bf{w}}} .
\end{split}
\end{equation}
\setcounter{equation}{\value{mytempeqncnt}}
%\vspace{-5pt}
\end{figure*}

\begin{figure*}[b!]
  \vspace{-0pt}
   \hrulefill
\setcounter{mytempeqncnt}{\value{equation}}
\setcounter{equation}{35}
\begin{equation}
\begin{split}{}
{\nabla _{\bf{x}}}{f_3}({\bf{w}},{\bf{x}}){\rm{ }} &= \sum\nolimits_{i = 1}^M {\frac{{{\beta _i}}}{{{K_i} + 1}}{K_i}\left( { - {{\bf{W}}_i}(2{\bf{C}}{{\bf{g}}_i} + 2{\bf{D}}{{\bf{q}}_i}) - {{\bf{S}}_i}(2{\bf{C}}{{\bf{q}}_i} - 2{\bf{D}}{{\bf{g}}_i})} \right)} \frac{{{\sigma ^2}}}{{{P_a}}}\left( {\frac{{{P_a}{{\left| {{{\bf{h}}_0}({\bf{x}}){\bf{w}}} \right|}^2}}}{{{\sigma ^2}{2^{{R_s}}}}} + \frac{1}{{{2^{{R_s}}}}} - 1} \right)\\
 & + \sum\nolimits_{i = 1}^M {\frac{{{\beta _i}}}{{{K_i} + 1}}\left( {{K_i}{{\left| {{{\overline {\bf{h}} }_i}({\bf{x}}){\bf{w}}} \right|}^2} + 1} \right)} {\beta _0}\frac{{ - {{\bf{W}}_0}(2{\bf{C}}{{\bf{g}}_0} + 2{\bf{D}}{{\bf{q}}_0}) - {{\bf{S}}_0}(2{\bf{C}}{{\bf{q}}_0} - 2{\bf{D}}{{\bf{g}}_0})}}{{{2^{{R_s}}}}}\\
 & - 2\kappa (\varepsilon )\sum\nolimits_{i = 1}^M {\frac{{{\beta _i}}}{{{K_i} + 1}}\left( {{K_i}{{\left| {{{\overline {\bf{h}} }_i}({\bf{x}}){\bf{w}}} \right|}^2} + 1} \right)} \sum\nolimits_{i = 1}^M {\frac{{{\beta _i}}}{{{K_i} + 1}}{K_i}\left( { - {{\bf{W}}_i}(2{\bf{C}}{{\bf{g}}_i} + 2{\bf{D}}{{\bf{q}}_i}) - {{\bf{S}}_i}(2{\bf{C}}{{\bf{q}}_i} - 2{\bf{D}}{{\bf{g}}_i})} \right)}  \\
 &- 2\rho (\varepsilon )\sum\nolimits_{i = 1}^M {{{\left( {\frac{{{\beta _i}}}{{{K_i} + 1}}} \right)}^2}{K_i}\left( { - {{\bf{W}}_i}(2{\bf{C}}{{\bf{g}}_i} + 2{\bf{D}}{{\bf{q}}_i}) - {{\bf{S}}_i}(2{\bf{C}}{{\bf{q}}_i} - 2{\bf{D}}{{\bf{g}}_i})} \right)}.
\end{split}
\end{equation}
\setcounter{equation}{\value{mytempeqncnt}}
%\vspace{-5pt}
\end{figure*}

We now present the key steps of applying the APGA for solving (P3.5) as shown in Algorithm 2, where given ${{\bf{x}}^{(t)}}$, Steps 3$-$9 aim to update ${{\bf{w}}^{(t)}}$ as ${{\bf{w}}^{(t + 1)}}$, and given ${{\bf{w}}^{(t + 1)}}$, Steps 10$-$16 aim to update ${{\bf{x}}^{(t)}}$ as ${{\bf{x}}^{(t + 1)}}$. These two procedures are implemented alternately until the objective converges to a stationary value. More specifically, Step 4 is used to update ${{\bf{w}}^{(t)}}$ without restraint, where $\delta $ is the step size and ${\nabla _{\bf{w}}}{f_3}({{\bf{w}}^{(t)}},{{\bf{x}}^{(t)}})$ is the gradient of ${f_3}({\bf{w}},{\bf{x}})$ w.r.t. ${\bf{w}}$ at $({{\bf{w}}^{(t)}},{{\bf{x}}^{(t)}})$. Step 5 is used to guarantee that ${{\bf{w}}^{(t + 1)}}$ always satisfies the constraint in (10b), i.e., $\left\| {{{\bf{w}}^{(t + 1)}}} \right\| = 1$. Subsequently, we need to further examine whether the inequality in Step 6 is established or not, where ${Q_\delta }({{\bf{w}}^{(t + 1)}},{{\bf{x}}^{(t)}};{{\bf{w}}^{(t)}})$ is a quadratic model of ${f_3}({\bf{w}},{\bf{x}})$ around ${{\bf{w}}^{(t)}}$ and can be expressed as \cite{RIS_MIMO_tsp}
\begin{equation}
\setcounter{equation}{33}
\begin{split}
&{Q_\delta }({{\bf{w}}^{(t + 1)}},{{\bf{x}}^{(t)}};{{\bf{w}}^{(t)}}) = {f_3}({{\bf{w}}^{(t)}},{{\bf{x}}^{(t)}})\\
 +& \left\langle {{\nabla _{\bf{w}}}{f_3}({{\bf{w}}^{(t)}},{{\bf{x}}^{(t)}}),{{\bf{w}}^{(t + 1)}} - {{\bf{w}}^{(t)}}} \right\rangle  - \frac{1}{\delta }{\left\| {{{\bf{w}}^{(t + 1)}} - {{\bf{w}}^{(t)}}} \right\|^2},
\end{split}
\end{equation}
with $\left\langle {{\bf{x}},{\bf{y}}} \right\rangle  = 2{\mathbb{R}}({{\bf{x}}^T}{\bf{y}})$. If the inequality in Step 6 is not feasible, we need to shrink the step size via Step 7 and then re-implement Steps 4$-$8 until the condition in Step 9 is established. Otherwise, if the inequality in Step 6 is feasible, the update for ${{\bf{w}}^{(t + 1)}}$ can be stopped and we can continue to update ${\bf{x}}$ in the next, for which the details are similar and thus are omitted here for brevity. The key point we should emphasize is that the update for ${{\bf{x}}^{(t + 1)}}$ via the projection function ${{\cal B}}\left\{  \cdot  \right\}$ in Step 12 is to guarantee that solutions for MAs' positions in each iteration always satisfy the constraint in (10c), where the projection function, based on the nearest-distance rule, can be easily derived as:
\begin{equation}
\begin{split}
{\cal B}\left\{ {{{\bf{x}}^{(t + 1)}},{\cal C}} \right\} \triangleright x_i^{(t + 1)} = \left\{ {\begin{array}{*{20}{c}}
{{F_i}}&{x_i^{(t + 1)} < {F_i}}\\
{x_i^{(t + 1)}}&{{F_i} \le x_i^{(t + 1)} \le {G_i}}\\
{{G_i}}&{x_i^{(t + 1)} > {G_i}}
\end{array}} \right..
\end{split}
\end{equation}

To successfully implement Algorithm 2, we have to further derive the expressions of ${\nabla _{\bf{w}}}{f_3}({\bf{w}},{\bf{x}})$ and ${\nabla _{\bf{x}}}{f_3}({\bf{w}},{\bf{x}})$, as shown in Corollary 1.

\textbf{Corollary 1:} ${\nabla _{{{\bf{w}}}}}{f_3}({\bf{w}},{\bf{x}})$ and ${\nabla _{{{\bf{x}}}}}{f_3}({\bf{w}},{\bf{x}})$ can be derived as in (35) and (36), where ${\bf{C}} = {\bf{u}}{{\bf{u}}^T} + {\bf{z}}{{\bf{z}}^T},{\bf{D}} = {\bf{u}}{{\bf{z}}^T} - {\bf{z}}{{\bf{u}}^T}$, ${\bf{u}}$ and ${\bf{z}}$ denote the real part and imaginary part of ${\bf{w}}$, respectively, i.e., ${\bf{w}} = {\bf{u}} + j{\bf{z}}$. In addition, for each $i = 0,1,...,M$, ${{\bf{g}}_i} = {\left[ {{g_{1,i}},{g_{2,i}},...,{g_{N,i}}} \right]^T}$, ${{\bf{q}}_i} = {\left[ {{q_{1,i}},{q_{2,i}},...,{q_{N,i}}} \right]^T}$, with
\begin{equation} \nonumber
\begin{split}
{g_{n,i}} = \cos \left( {\frac{{2\pi }}{\lambda }{x_n}\sin {\theta _i}} \right),{q_{n,i}} =  - \sin \left( {\frac{{2\pi }}{\lambda }{x_n}\sin {\theta _i}} \right),
\end{split}
\end{equation}
and
\begin{equation} \nonumber
\begin{split}
{{\bf{W}}_i} =& {\rm{diag}}\left( {\left\{ {\frac{{2\pi }}{\lambda }\sin {\theta _i}\sin \left( {\frac{{2\pi }}{\lambda }{x_n}\sin {\theta _i}} \right)} \right\}_{n = 1}^N} \right),\\
{{\bf{S}}_i} =& {\rm{diag}}\left( {\left\{ {\frac{{2\pi }}{\lambda }\sin {\theta _i}\cos \left( {\frac{{2\pi }}{\lambda }{x_n}\sin {\theta _i}} \right)} \right\}_{n = 1}^N} \right).
\end{split}
\end{equation}

\begin{proof}
Corollary 1 can be proved by resorting to the differentiation method and further noting that:
 \begin{equation} \nonumber
\begin{split}
{\nabla _{\bf{w}}}{\left| {{{\overline {\bf{h}} }_i}({\bf{x}}){\bf{w}}} \right|^2} =& \overline {\bf{h}} _i^H({\bf{x}}){\overline {\bf{h}} _i}({\bf{x}}){\bf{w}}, i = 1,2,...,M,\\
{\nabla _{\bf{w}}}{\left| {{{\bf{h}}_0}({\bf{x}}){\bf{w}}} \right|^2} =& {\bf{h}}_0^H{{\bf{h}}_0}{\bf{w}},\\
\end{split}
\end{equation}
and
 \begin{equation} \nonumber
\begin{split}
{\nabla _{\bf{x}}}{\left| {{{\overline {\bf{h}} }_i}({\bf{x}}){\bf{w}}} \right|^2} \mathop  = \limits^{(b)} &  - {{\bf{W}}_i}(2{\bf{C}}{{\bf{g}}_i} + 2{\bf{D}}{{\bf{q}}_i})\\
 &- {{\bf{S}}_i}(2{\bf{C}}{{\bf{q}}_i} - 2{\bf{D}}{{\bf{g}}_i}),i = 1,2,...M,\\
{\nabla _{\bf{x}}}{\left| {{{\bf{h}}_0}({\bf{x}}){\bf{w}}} \right|^2} \mathop  = \limits^{(c)} & {\beta _0}( - {{\bf{W}}_0}(2{\bf{C}}{{\bf{g}}_0} + 2{\bf{D}}{{\bf{q}}_0})\\
 -& {{\bf{S}}_0}(2{\bf{C}}{{\bf{q}}_0} - 2{\bf{D}}{{\bf{g}}_0})),
\end{split}
\end{equation}
where equalities $\mathop  = \limits^{(b)} $ and $\mathop  = \limits^{(c)} $ are established based on conclusions of our previous work \cite{Guojie_SPL}.
\end{proof}

\textit{Convergence analysis:} According to \cite{RIS_MIMO_tsp}, since in each iteration, the suitable step size is chosen based on the backtracking line search to strictly guarantee that the objective is no smaller than the quadratic model, the objective will be strictly increasing w.r.t. the number of iterations. Moreover, ${f_3}({\bf{w}},{\bf{x}})$ has its upper bound since the values of ${{{\left| {{{\bf{h}}_0}({\bf{x}}){\bf{w}}} \right|}^2}}$ and ${\left| {{{\overline {\bf{h}} }_i}({\bf{x}}){\bf{w}}} \right|^2}$, $i = 1,2,...,M$, are all bounded given the constraints in (10b) and (10c). Therefore, the objective value will converge to a stationary point.

\textit{The complexity of Algorithm 2:} To simplify the analysis while still keeping a good approximation to the complexity of Algorithm 2, we focus on the number of complex multiplications required per iteration. Based on this rule and note that given ${{\bf{w}}^{(t)}}$ and ${{\bf{x}}^{(t)}}$, ${\nabla _{\bf{w}}}{f_3}({{\bf{w}}^{(t)}},{{\bf{x}}^{(t)}})$ in Step 4 is computed only once, and the complexity mainly comes from computing $\overline {\bf{h}} _i^H({{\bf{x}}^{(t)}}){\overline {\bf{h}} _i}({{\bf{x}}^{(t)}}){{\bf{w}}^{(t)}}$, $i = 1,2,...,M$, and ${\bf{h}}_0^H({{\bf{x}}^{(t)}}){{\bf{h}}_0}({{\bf{x}}^{(t)}}){{\bf{w}}^{(t)}}$, which is about ${\cal O}({N^2}M)$. While ${f_3}({{\bf{w}}^{(t + 1)}},{{\bf{x}}^{(t)}})$ and ${Q_\delta }({{\bf{w}}^{(t + 1)}},{{\bf{x}}^{(t)}};{{\bf{w}}^{(t)}})$ in Step 6 are computed frequently with the update of the step size in Step 7 (which results in the update of ${{\bf{w}}^{(t + 1)}}$ in Step 5). Focusing on the expression of ${f_3}({{\bf{w}}^{(t + 1)}},{{\bf{x}}^{(t)}})$, since the complexity of computing ${\left| {{{\bf{h}}_0}({{\bf{x}}^{(t)}}){{\bf{w}}^{(t + 1)}}} \right|^2}$ and ${\left| {{{\overline {\bf{h}} }_i}({{\bf{x}}^{(t)}}){{\bf{w}}^{(t + 1)}}} \right|^2}$, $i = 1,2,...,M$, is about ${\cal O}(N)$, the complexity of computing ${f_3}({{\bf{w}}^{(t + 1)}},{{\bf{x}}^{(t)}})$ would be ${{\cal O}}\left( {{N}M} \right)$. Also, focusing on the expression of ${Q_\delta }({{\bf{w}}^{(t + 1)}},{{\bf{x}}^{(t)}};{{\bf{w}}^{(t)}})$, the complexity of obtaining $\left\langle {{\nabla _{\bf{w}}}{f_3}({{\bf{w}}^{(t)}},{{\bf{x}}^{(t)}}),{{\bf{w}}^{(t + 1)}} - {{\bf{w}}^{(t)}}} \right\rangle $ is only ${\cal O}(N)$ since ${\nabla _{\bf{w}}}{f_3}({{\bf{w}}^{(t)}},{{\bf{x}}^{(t)}})$ is already known. Then, further denote the number of iterations to obtain a feasible step size for updating ${\bf{w}}$ as ${{I_{\delta ,{\bf{w}}}}}$, the complexity of Steps 3$-$9 is about ${\cal O}({N^2}M + {I_{\delta ,{\bf{w}}}}(NM + N))$. Using the same analysis, the complexity of Steps 10$-$16 can be derived as ${\cal O}({N^2}M + {I_{\delta ,{\bf{x}}}}(NM + N))$, where ${{I_{\delta ,{\bf{x}}}}}$ is the number of iterations to obtain a feasible step size for updating ${\bf{x}}$. Hence, the total complexity of Algorithm 2 is about ${\cal O}\left( {{I_{{\rm{out}}}}({N^2}M + ({I_{\delta ,{\bf{w}}}} + {I_{\delta ,{\bf{x}}}})(NM + N))} \right)$, with ${{I_{{\rm{out}}}}}$ denoting the number of iterations for alternately optimizing ${\bf{w}}$ and ${\bf{x}}$.

\textit{The total complexity of Algorithm 1:} Armed with the above analysis, we can finally compute the total complexity for solving the initial problem (P3.2) as ${\cal O}\left( {\log (1/\tau ){I_{{\rm{out}}}}({N^2}M + ({I_{\delta ,{\bf{w}}}} + {I_{\delta ,{\bf{x}}}})(NM + N))} \right)$.

%\begin{figure}
%\centering
%\includegraphics[width=7cm]{linear_model.eps}
%\captionsetup{font=small}
%\caption{${\gamma ^{ - 1}}\left( {\varepsilon ,y} \right)$ and the proposed linear approximation model w.r.t. $y$ under different $\varepsilon $.} \label{fig:Fig1}
%%\vspace{-5pt}
%\end{figure}

%\begin{figure}
%\centering
%\includegraphics[width=7cm]{linear_model.eps}
%\captionsetup{font=small}
%\caption{${\gamma ^{ - 1}}\left( {\varepsilon ,y} \right)$ and the proposed linear approximation model w.r.t. $y$ under different $\varepsilon $.} \label{fig:Fig1}
%%\vspace{-5pt}
%\end{figure}

%\begin{figure}
%\centering
%\includegraphics[width=7cm]{linear_model.eps}
%\captionsetup{font=small}
%\caption{${\gamma ^{ - 1}}\left( {\varepsilon ,y} \right)$ and the proposed linear approximation model w.r.t. $y$ under different $\varepsilon $.} \label{fig:Fig1}
%%\vspace{-5pt}
%\end{figure}
%
%\begin{figure}
%\centering
%\includegraphics[width=7cm]{linear_model.eps}
%\captionsetup{font=small}
%\caption{${\gamma ^{ - 1}}\left( {\varepsilon ,y} \right)$ and the proposed linear approximation model w.r.t. $y$ under different $\varepsilon $.} \label{fig:Fig1}
%%\vspace{-5pt}
%\end{figure}
%
%\begin{figure}
%\centering
%\includegraphics[width=7cm]{linear_model.eps}
%\captionsetup{font=small}
%\caption{${\gamma ^{ - 1}}\left( {\varepsilon ,y} \right)$ and the proposed linear approximation model w.r.t. $y$ under different $\varepsilon $.} \label{fig:Fig1}
%%\vspace{-5pt}
%\end{figure}

  \begin{algorithm}
\caption{The PGD for Solving (P4)}
  \begin{algorithmic}[1]

\State \textbf{{Input:}} ${{\bf{x}}^{(0)}} \in {\cal C}$, $t \leftarrow 0$, $\delta  > 0$ and $\chi  < 1$.

%\State \textbf{Initialize:} ${\varepsilon _1} = 0.5$, ${\varepsilon _L} = 0$ and ${\varepsilon _R} = 1$.

%\State \quad For $k = 1:1:M/2 - 1$:

\State \ \ \textbf{Repeat:} \quad \quad $\% \% \% \% \%$  \textit{The steps for updating} ${\bf{x}}$

\State \quad \ ${{\bf{x}}^{(t + 1)}} = {{\bf{x}}^{(t)}} - \delta {\nabla _{\bf{x}}}\Theta ({{\bf{x}}^{(t)}})$;

\State \quad \ ${{\bf{x}}^{(t + 1)}} = {\cal B}\left\{ {{{\bf{x}}^{(t + 1)}},{\cal C}} \right\}$;

\State \quad \ {\textbf{if}} $\Theta ({{\bf{x}}^{(t + 1)}}) < {Q_\delta }({{\bf{x}}^{(t + 1)}};{{\bf{x}}^{(t)}})$, {\textbf{then}}

\State \quad \quad \ $\delta  \leftarrow \chi \delta $,

\State \quad \ \textbf{end}

\State \ \ \textbf{Until:} $\Theta ({{\bf{x}}^{(t + 1)}}) \ge {Q_\delta }({{\bf{x}}^{(t + 1)}};{{\bf{x}}^{(t)}})$.

\State \quad $t \leftarrow t + 1$;

\State \textbf{Until:} $\Theta ({{\bf{x}}^{(t)}})$ converges to a stationary point.
  \end{algorithmic}
\end{algorithm}

\section{A Competitive Benchmark}
In this section, we further present a competitive benchmark for handling (P3). This benchmark still employs the MA technique to reap the additional spatial DoF and concurrently adopts the ZF-based transmit beamforming due to that: i) it can effectively decrease the averaged signal power leaked to eavesdroppers; ii) it enjoys low implementation complexities especially when the number of antennas at Alice is larger.

Specifically, employing the ZF-based beamforming ${{{\bf{w}}_{{\rm{ZF}}}}}$ and without eavesdroppers' instantaneous CSI, Alice aims to force the receiving power gain at $\left\{ {{\rm{Ev}}{{\rm{e}}_i}} \right\}_{i = 1}^M$ via the LoS paths to zero, i.e., ${\left| {{{\overline {\bf{h}} }_i}({\bf{x}}){{\bf{w}}_{{\rm{ZF}}}}} \right|^2} = 0$, $i = 1,2,...,M$, and concurrently maximize the receiving power gain at Bob. Note that this operation requires $N \ge M + 1$.

Given positions of antennas at Alice as ${\bf{x}}$, ${{{\bf{w}}_{{\rm{ZF}}}}}$ can be expressed as ${{\bf{w}}_{{\rm{ZF}}}}({\bf{x}}) = {\bf{w}}({\bf{x}})/\left\| {{\bf{w}}({\bf{x}})} \right\|$ \cite{Zhulipeng_CL}, with
 \begin{equation}
\setcounter{equation}{37}
\begin{split}
{\bf{w}}({\bf{x}}) =& \left[ {{{\bf{I}}_N} - {\bf{H}}({\bf{x}}){{\left( {{{\bf{H}}^H}({\bf{x}}){\bf{H}}({\bf{x}})} \right)}^{ - 1}}{{\bf{H}}^H}({\bf{x}})} \right]{\bf{h}}_0^H({\bf{x}}),\\
{\bf{H}}({\bf{x}}) =& {[{\overline {\bf{h}} _1}({\bf{x}});{\overline {\bf{h}} _2}({\bf{x}});...;{\overline {\bf{h}} _M}({\bf{x}})]^H} \in {{\mathbb{C}}^{N \times M}},
\end{split}
\end{equation}
 and ${\left(  \cdot  \right)^{ - 1}}$ denotes the inverse operation. Based on the known ${{{\bf{w}}_{{\rm{ZF}}}}}$, the objective of (P3) can be simplified as $\gamma \left( {{\psi _1},{\psi _2}\left( {\frac{{{P_a}{{\left| {{{\bf{h}}_0}({\bf{x}}){{\bf{w}}_{{\rm{ZF}}}}({\bf{x}})} \right|}^2}}}{{{\sigma ^2}{2^{{R_s}}}}} + \frac{1}{{{2^{{R_s}}}}} - 1} \right)} \right)$,
%  \begin{equation}\nonumber
%\begin{split}
%\gamma \left( {{\psi _1},{\psi _2}\left( {\frac{{{P_a}{{\left| {{{\bf{h}}_0}({\bf{x}}){{\bf{w}}_{{\rm{ZF}}}}({\bf{x}})} \right|}^2}}}{{{\sigma ^2}{2^{{R_s}}}}} + \frac{1}{{{2^{{R_s}}}}} - 1} \right)} \right),
%\end{split}
%\end{equation}
 where ${\psi _1} = \frac{{{{\left( {\sum\nolimits_{i = 1}^M {\frac{{{\beta _i}}}{{{K_i} + 1}}} } \right)}^2}}}{{\sum\nolimits_{i = 1}^M {{{\left( {\frac{{{\beta _i}}}{{{K_i} + 1}}} \right)}^2}} }}$ and ${\psi _2} = \frac{{\sum\nolimits_{i = 1}^M {\frac{{{\beta _i}}}{{{K_i} + 1}}} }}{{\sum\nolimits_{i = 1}^M {{{\left( {\frac{{{\beta _i}}}{{{K_i} + 1}}} \right)}^2}} }}\frac{{{\sigma ^2}}}{{{P_a}}}$.

Observing the above simplified objective, to minimize the secrecy outage probability, it is obviously equivalent to maximizing ${{{\left| {{{\bf{h}}_0}({\bf{x}}){{\bf{w}}_{{\rm{ZF}}}}({\bf{x}})} \right|}^2}}$, which can be further expanded as
  \begin{equation}
\begin{split}
&{\left| {{{\bf{h}}_0}({\bf{x}}){{\bf{w}}_{{\rm{ZF}}}}({\bf{x}})} \right|^2}\\
 =& {\left| {{{\bf{h}}_0}({\bf{x}})\left[ {{{\bf{I}}_N} - {\bf{H}}({\bf{x}}){{\left( {{{\bf{H}}^H}({\bf{x}}){\bf{H}}({\bf{x}})} \right)}^{ - 1}}{{\bf{H}}^H}({\bf{x}})} \right]{\bf{h}}_0^H({\bf{x}})} \right|}\\
 =& {\beta _0}N - \underbrace {{{\bf{h}}_0}({\bf{x}}){\bf{H}}({\bf{x}}){{\left( {{{\bf{H}}^H}({\bf{x}}){\bf{H}}({\bf{x}})} \right)}^{ - 1}}{{\bf{H}}^H}({\bf{x}}){\bf{h}}_0^H({\bf{x}})}_{\Theta ({\bf{x}})}.
\end{split}
\end{equation}

Therefore, to maximize ${{{\left| {{{\bf{h}}_0}({\bf{x}}){{\bf{w}}_{{\rm{ZF}}}}({\bf{x}})} \right|}^2}}$, it is equivalent to minimizing ${\Theta ({\bf{x}})}$, leading to the optimization problem as
\begin{align}
&({\rm{P4}}):{\rm{  }}\mathop {\min }\limits_{{{\bf{x}}}} \ {\Theta ({\bf{x}})}  \tag{${\rm{39a}}$}\\
{\rm{              }}&\ {\rm{s.t.}} \quad (10{\rm{c}}).\tag{${\rm{39b}}$}
%& \ \ \ \quad {1_{\mathbb{C}}} = 1,\ {\rm{if}} \ \\
%&0 < {Q_E} \le \max \left( {{Q_{th}}/\left( {\frac{1}{{{\lambda _{EE}}}} + \frac{{{\beta %^r}N}}{{{\lambda _{RE}}{\lambda _{ER}}}}} \right),{Q_{E,\max }}} \right)\tag{${\rm{11d}}$}.
 \end{align}

 \begin{figure*}[b!]
  \vspace{-0pt}
   \hrulefill
\setcounter{mytempeqncnt}{\value{equation}}
\setcounter{equation}{41}
\begin{equation}
\begin{split}{}
\frac{{\partial \Theta ({\bf{x}})}}{{\partial {x_n}}} =& \frac{{\partial {\bf{h}}({\bf{x}})}}{{\partial {x_n}}}{\left( {{{\bf{H}}^H}({\bf{x}}){\bf{H}}({\bf{x}})} \right)^{ - 1}}{{\bf{h}}^H}({\bf{x}}) + {\bf{h}}({\bf{x}}){\left( {{{\bf{H}}^H}({\bf{x}}){\bf{H}}({\bf{x}})} \right)^{ - 1}}\frac{{\partial {{\bf{h}}^H}({\bf{x}})}}{{\partial {x_n}}}\\
 &+ {\bf{h}}({\bf{x}})\left[ { - {{\left( {{{\bf{H}}^H}({\bf{x}}){\bf{H}}({\bf{x}})} \right)}^{ - 1}}\left[ {\frac{{\partial {{\bf{H}}^H}({\bf{x}})}}{{\partial {x_n}}}{\bf{H}}({\bf{x}}) + {{\bf{H}}^H}({\bf{x}})\frac{{\partial {\bf{H}}({\bf{x}})}}{{\partial {x_n}}}} \right]{{\left( {{{\bf{H}}^H}({\bf{x}}){\bf{H}}({\bf{x}})} \right)}^{ - 1}}} \right]{{\bf{h}}^H}({\bf{x}}).
\end{split}
\end{equation}
\setcounter{equation}{\value{mytempeqncnt}}
%\vspace{-5pt}
\end{figure*}

Since ${\Theta ({\bf{x}})}$ is neither convex or concave w.r.t. ${\bf{x}}$, we now exploit the PGD to solve (P4), the key steps of which are similar to those shown in Algorithm 2. Thus, we here provide a brief description as presented in Algorithm 3, where ${\nabla _{\bf{x}}}\Theta ({{\bf{x}}^{(t)}})$ is the gradient of $\Theta ({\bf{x}})$ w.r.t. ${\bf{x}}$ at ${{\bf{x}}^{(t)}}$, and ${Q_\delta }({{\bf{x}}^{(t + 1)}};{{\bf{x}}^{(t)}})$ is a quadratic model of $\Theta ({\bf{x}})$ around ${{\bf{x}}^{(t)}}$ and can be expressed as
  \begin{equation}
  \setcounter{equation}{40}
\begin{split}
&{Q_\delta }({{\bf{x}}^{(t + 1)}};{{\bf{x}}^{(t)}}) = \Theta ({{\bf{x}}^{(t)}})\\
 +& \left\langle {{\nabla _{\bf{x}}}\Theta ({{\bf{x}}^{(t)}}),{{\bf{x}}^{(t + 1)}} - {{\bf{x}}^{(t)}}} \right\rangle  - \frac{1}{\delta }{\left\| {{{\bf{x}}^{(t + 1)}} - {{\bf{x}}^{(t)}}} \right\|^2}.
\end{split}
\end{equation}

%\begin{figure*}[[b!]
%  \vspace{-0pt}
%   \hrulefill
%\setcounter{mytempeqncnt}{\value{equation}}
%\setcounter{equation}{42}
%\begin{equation}
%\begin{split}{}
%\frac{{\partial {\bf{h}}({\bf{x}})}}{{\partial {x_n}}} =& \sqrt {{\beta _0}} \frac{{2\pi }}{\lambda }\left[ {(\sin {\theta _0} - \sin {\theta _1}){e^{j\left[ {\frac{{2\pi }}{\lambda }{x_n}(\sin {\theta _0} - \sin {\theta _1}) + \frac{\pi }{2}} \right]}},...,(\sin {\theta _0} - \sin {\theta _M}){e^{j\left[ {\frac{{2\pi }}{\lambda }{x_n}(\sin {\theta _0} - \sin {\theta _M}) + \frac{\pi }{2}} \right]}}} \right].
%\end{split}
%\end{equation}
%\setcounter{equation}{\value{mytempeqncnt}}
%%\vspace{-5pt}
%\end{figure*}

We now focus on deriving the closed-form ${\nabla _{\bf{x}}}\Theta ({\bf{x}})$ as follows. Specifically, based on the known expressions of ${{\bf{h}}_0}({\bf{x}})$ and ${\bf{H}}({\bf{x}})$ shown in (37), we denote
  \begin{equation}
\begin{split}
{\bf{h}}({\bf{x}})  \buildrel \Delta \over = & {{\bf{h}}_0}({\bf{x}}){\bf{H}}({\bf{x}})\\
 = &\sqrt {{\beta _0}} {\left[ {\begin{array}{*{20}{c}}
{\sum\nolimits_{i = 1}^N {{e^{j\frac{{2\pi }}{\lambda }{x_i}(\sin {\theta _0} - \sin {\theta _1})}}} }\\
 \vdots \\
{\sum\nolimits_{i = 1}^N {{e^{j\frac{{2\pi }}{\lambda }{x_i}(\sin {\theta _0} - \sin {\theta _M})}}} }
\end{array}} \right]^T},
\end{split}
\end{equation}
and thus $\Theta ({\bf{x}}) = {\bf{h}}({\bf{x}}){\left( {{{\bf{H}}^H}({\bf{x}}){\bf{H}}({\bf{x}})} \right)^{ - 1}}{{\bf{h}}^H}({\bf{x}})$.

\textbf{Corollary 2:} ${\nabla _{\bf{x}}}\Theta ({\bf{x}}) = {\left[ {\frac{{\partial \Theta ({\bf{x}})}}{{\partial {x_1}}},\frac{{\partial \Theta ({\bf{x}})}}{{\partial {x_2}}},...,\frac{{\partial \Theta ({\bf{x}})}}{{\partial {x_N}}}} \right]^T}$, where ${\frac{{\partial \Theta ({\bf{x}})}}{{\partial {x_n}}}}$, $n = 1,2,...,N$, is derived in (42).
\begin{proof}
Note that
\begin{equation}
 \setcounter{equation}{43}
\begin{split}
\frac{{\partial \Theta ({\bf{x}})}}{{\partial {x_n}}} =& \frac{{\partial {\bf{h}}({\bf{x}})}}{{\partial {x_n}}}{\left( {{{\bf{H}}^H}({\bf{x}}){\bf{H}}({\bf{x}})} \right)^{ - 1}}{{\bf{h}}^H}({\bf{x}})\\
 &+ {\bf{h}}({\bf{x}})\frac{{\partial {{\left( {{{\bf{H}}^H}({\bf{x}}){\bf{H}}({\bf{x}})} \right)}^{ - 1}}}}{{\partial {x_n}}}{{\bf{h}}^H}({\bf{x}})\\
 &+ {\bf{h}}({\bf{x}}){\left( {{{\bf{H}}^H}({\bf{x}}){\bf{H}}({\bf{x}})} \right)^{ - 1}}\frac{{\partial {{\bf{h}}^H}({\bf{x}})}}{{\partial {x_n}}},
\end{split}
\end{equation}
and since $d{{\bf{A}}^{ - 1}} =  - {{\bf{A}}^{ - 1}}(d{\bf{A}}){{\bf{A}}^{ - 1}}$, we can obtain
  \begin{equation}
\begin{split}
&\frac{{\partial {{\left( {{{\bf{H}}^H}({\bf{x}}){\bf{H}}({\bf{x}})} \right)}^{ - 1}}}}{{\partial {x_n}}}\\
 = & - {\left( {{{\bf{H}}^H}({\bf{x}}){\bf{H}}({\bf{x}})} \right)^{ - 1}}\frac{{\partial \left( {{{\bf{H}}^H}({\bf{x}}){\bf{H}}({\bf{x}})} \right)}}{{\partial {x_n}}}{\left( {{{\bf{H}}^H}({\bf{x}}){\bf{H}}({\bf{x}})} \right)^{ - 1}},
\end{split}
\end{equation}
where
  \begin{equation}
\begin{split}
\frac{{\partial \left( {{{\bf{H}}^H}({\bf{x}}){\bf{H}}({\bf{x}})} \right)}}{{\partial {x_n}}} = \frac{{\partial {{\bf{H}}^H}({\bf{x}})}}{{\partial {x_n}}}{\bf{H}}({\bf{x}}) + {{\bf{H}}^H}({\bf{x}})\frac{{\partial {\bf{H}}({\bf{x}})}}{{\partial {x_n}}}.
\end{split}
\end{equation}
Substituting (45) into (44) and further substituting (44) into (43), we can obtain the result in (42). This completes the proof.
\end{proof}

Given ${\bf{x}}$, to compute the value of $\frac{{\partial \Theta ({\bf{x}})}}{{\partial {x_n}}}$, we need to derive the expressions of $\frac{{\partial {\bf{h}}({\bf{x}})}}{{\partial {x_n}}}$ ($\frac{{\partial {{\bf{h}}^H}({\bf{x}})}}{{\partial {x_n}}} = {\left[ {\frac{{\partial {\bf{h}}({\bf{x}})}}{{\partial {x_n}}}} \right]^H}$) and ${\frac{{\partial {\bf{H}}({\bf{x}})}}{{\partial {x_n}}}}$ ($\frac{{\partial {{\bf{H}}^H}({\bf{x}})}}{{\partial {x_n}}} = {\left[ {\frac{{\partial {\bf{H}}({\bf{x}})}}{{\partial {x_n}}}} \right]^H}$) as
\begin{equation}
  \begin{split}
&\frac{{\partial {\bf{h}}({\bf{x}})}}{{\partial {x_n}}} \\
 =& \sqrt {{\beta _0}} \frac{{2\pi }}{\lambda }{\left[ {\begin{array}{*{20}{c}}
{(\sin {\theta _0} - \sin {\theta _1}){e^{j\left[ {\frac{{2\pi }}{\lambda }{x_n}(\sin {\theta _0} - \sin {\theta _1}) + \frac{\pi }{2}} \right]}}}\\
 \vdots \\
{(\sin {\theta _0} - \sin {\theta _M}){e^{j\left[ {\frac{{2\pi }}{\lambda }{x_n}(\sin {\theta _0} - \sin {\theta _M}) + \frac{\pi }{2}} \right]}}}
\end{array}} \right]^T},
\end{split}
\end{equation}
and
\begin{equation}
\begin{split}
\frac{{\partial {\bf{H}}({\bf{x}})}}{{\partial {x_n}}} = \frac{{2\pi }}{\lambda }\left[ {\begin{array}{*{20}{c}}
{{{\bf{0}}^{1 \times M}}}\\
{...}\\
{{\bf{b}}({x_n})}\\
{...}\\
{{{\bf{0}}^{1 \times M}}}
\end{array}} \right],
\end{split}
\end{equation}
 with
  \begin{equation}
\begin{split}
{\bf{b}}({x_n}) = {\left[ {\begin{array}{*{20}{c}}
{\sin {\theta _1}{e^{ - j\left( {\frac{{2\pi }}{\lambda }{x_n}\sin {\theta _1} + \frac{\pi }{2}} \right)}}}\\
 \vdots \\
{\sin {\theta _M}{e^{ - j\left( {\frac{{2\pi }}{\lambda }{x_n}\sin {\theta _M} + \frac{\pi }{2}} \right)}}}
\end{array}} \right]^T}.
\end{split}
\end{equation}

\textit{Convergence analysis:} Similarly, since the suitable step size is chosen based on the backtracking line search to guarantee that the objective is not larger than the quadratic model, the objective will be strictly decreasing w.r.t. the number of iterations. In addition, obviously $\Theta ({\bf{x}})$ has its lower bound as zero, at which the maximum receiving power gain is achieved at Bob, i.e., ${\left| {{{\bf{h}}_0}({\bf{x}}){{\bf{w}}_{{\rm{ZF}}}}({\bf{x}})} \right|^2} = {\beta _0}N - \Theta ({\bf{x}}) = {\beta _0}N$.  Based on the above analysis, the objective value of (P4) will converge to a stationary point via Algorithm 3.

\textit{The complexity of Algorithm 3:} We first analyze the complexity of computing $\frac{{\partial \Theta ({\bf{x}})}}{{\partial {x_n}}}$, $n = 1,2,...,N$. Specifically, note that computing ${{\bf{H}}^H}({\bf{x}}){\bf{H}}({\bf{x}})$,
$\frac{{\partial {{\bf{H}}^H}({\bf{x}})}}{{\partial {x_n}}}{\bf{H}}({\bf{x}})$ and $\frac{{\partial {\bf{H}}({\bf{x}})}}{{\partial {x_n}}}{{\bf{H}}^H}({\bf{x}})$ requires the complexities of ${{\cal O}}(N{M^2})$, ${{\cal O}}(N{M^2})$ and ${{\cal O}}(M{N^2})$, respectively. Further, for a matrix ${\bf{A}} \in {{\mathbb{C}}^{M \times M}}$, computing ${{\bf{A}}^{ - 1}}$ requires the complexity of ${{\cal O}}({M^3})$. Based on the above analysis, the complexity of computing $\frac{{\partial {{\left( {{{\bf{H}}^H}({\bf{x}}){\bf{H}}({\bf{x}})} \right)}^{ - 1}}}}{{\partial {x_n}}}$ is about ${{\cal O}}(N{M^2} + {M^3})$, according to which it is easy to derive the complexity of computing $\frac{{\partial \Theta ({\bf{x}})}}{{\partial {x_n}}}$ as ${{\cal O}}(N{M^2} + {M^3})$. Therefore, obtaining ${\nabla _{\bf{x}}}\Theta ({\bf{x}})$ requires the complexity of ${{\cal O}}({N^2}{M^2} + N{M^3})$. Next, using the similar analysis method in the above for handing Algorithm 2, we can directly derive the total complexity for solving (P4) as ${{\cal O}}\left( {{I_{{\rm{out}}}}({N^2}{M^2} + N{M^3} + {I_{\delta ,{\bf{x}}}}(N{M^2} + {M^3}))} \right)$, where ${I_{\delta ,{\bf{x}}}}$ is the number of iterations to obtain a feasible step size and ${I_{{\rm{out}}}}$ is the number of iterations until the objective converges to a stationary value.

\section{Simulation Results}
In this section, we provide numerical results to demonstrate the effectiveness of our proposed algorithms for minimizing the secrecy outage probability. Unless otherwise stated, the minimum distance between any two MAs is set as ${d_{\min }} = \lambda /2$, where $\lambda $ is set as one for simplification and without affecting the obtained results, the Rician K-factors for all eavesdropping channels are set to be the same , i.e., $\left\{ {{K_i}} \right\}_{i = 1}^M = K$, and the targeted PLS coding rate of Alice's encoder is set as $R_s = 3$ bits/s/Hz. In addition, we set the large-scale fading power of the main channel and the eavesdropping channels as $\left\{ {{\beta _i}} \right\}_{i = 0}^M = 1$. Here, such power gains are normalized over the receiver noise power so that we can conveniently set the noise power as ${\sigma ^2} = 1$.

\begin{figure}
\centering
\includegraphics[width=8.4cm]{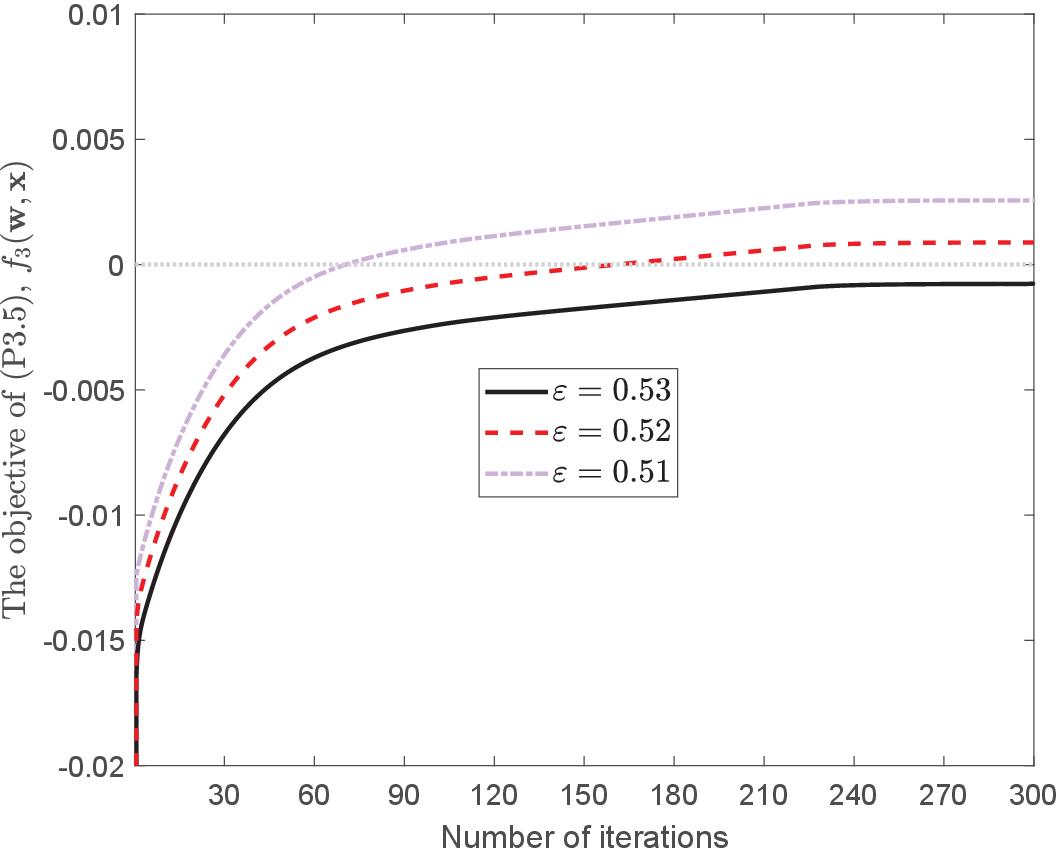}
\captionsetup{font=small}
\caption{Convergence behavior of the proposed APGA for handing (P3.5), where ${P_a} = 25$ dB, $N = 5$, $M = 1$, $L = 4$, $K = 4$, ${\theta _0} = \frac{\pi }{4}$ and ${\theta _1} = \frac{{1.1\pi }}{4}$.} \label{fig:Fig1}
%\vspace{-5pt}
\end{figure}

\begin{figure}
\centering
\includegraphics[width=8cm]{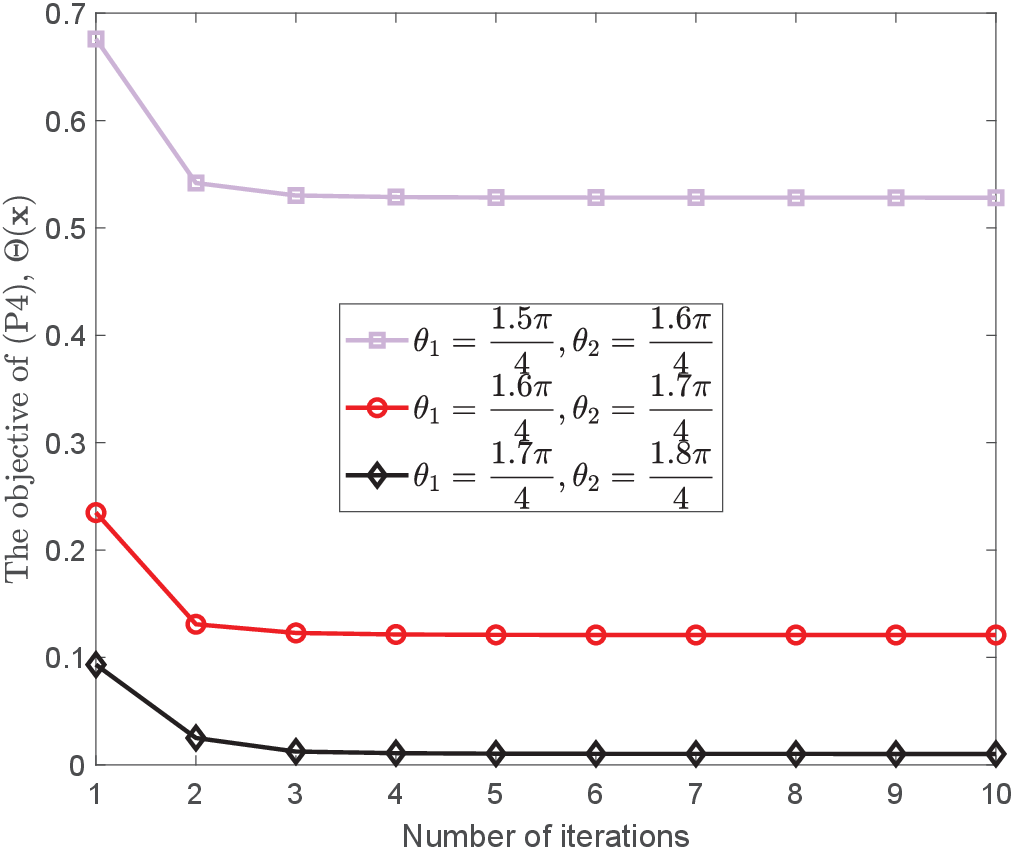}
\captionsetup{font=small}
\caption{Convergence behavior of the proposed PGD for handling (P4), where ${P_a} = 15$ dB, $N = 5$, $M = 2$, $L = 4$, $K = 4$ and ${\theta _0} = \frac{\pi }{4}$.} \label{fig:Fig1}
%\vspace{-5pt}
\end{figure}

\subsection{Convergence Analysis of the APGA and the PGD for Handing (P3.5) and (P4)}
We first illustrate the convergence behavior of the proposed APGA for solving (P3.5) as shown Fig. 6, from which it is observed that: i) given each $\varepsilon  = 0.51,0.52,0.53$, the objective of (P3.5) always converges to a stationary value after hundreds of iterations, thus verifying the effectiveness of our proposed algorithm; ii) as $\varepsilon$ increases, the stationary value of the objective of (P3.5) will decrease accordingly, and this behavior can be explained based on (30). That is to say, as $\varepsilon$ increases, $\kappa (\varepsilon ){f_1}({\bf{w}},{\bf{x}}) + \rho (\varepsilon )$ will increase by observing Fig. 3. Thus, ${f_2}({\bf{w}},{\bf{x}}) - \kappa (\varepsilon ){f_1}({\bf{w}},{\bf{x}}) - \rho (\varepsilon )$ will inevitably decrease w.r.t. $\varepsilon $; iii) note that Algorithm 1 aims to determine the maximum value of $\varepsilon $ that makes the optimal output of (P3.5) not smaller than zero. Based on this conclusion, it is easy to determine that the optimal $\varepsilon $ of (P3.2) is 0.52. Then, the obtained secrecy outage probability in this case is ${P_{{\rm{out}}}} = 1 - \varepsilon  = 0.48$.

We now illustrate the convergence behavior of the proposed PGD for solving (P4) as shown in Fig. 7, from which it is observed that: i) given different combinations of ${\theta _1}$ and ${\theta _2}$, the objective of (P4) always converges to a stationary value after about five iterations, which indicates that Algorithm 3 enjoys a low complexity; ii) the stationary objective of (P4), i.e., $\Theta ({\bf{x}})$, becomes smaller when ${\theta _1}$ and ${\theta _2}$ become larger. For instance, when ${\theta _1} = \frac{{1.7\pi }}{4}$ and ${\theta _2} = \frac{{1.8\pi }}{4}$, the stationary value of $\Theta ({\bf{x}})$ approaches zero; while when ${\theta _1} = \frac{{1.5\pi }}{4}$ and ${\theta _2} = \frac{{1.6\pi }}{4}$, the stationary value of $\Theta ({\bf{x}})$ is even greater than 0.5. This phenomenon is due to that in this case (${\theta _0} = \frac{\pi }{4}$), when ${\theta _1}$ and ${\theta _2}$ increases, the correlation between the main channel and the wiretap channel becomes smaller, and thus $\Theta ({\bf{x}})$ should become smaller such that the receiving power gain at Bob via the ZF-based beamforming, i.e., ${\beta _0}N - \Theta ({\bf{x}})$, becomes larger.

\subsection{Performance Comparison}
We now formally present the achievable performance of our proposed algorithms. For simplification, we denote the schemes designed in Algorithm 1 and Algorithm 3 as MA+OB and MA+ZF, respectively. In addition, for comprehensive comparisons, we further consider the following five benchmarks:
\begin{itemize}
\item RAP+OB: Randomly generate ${\bf{x}}$ satisfying (10c) for 100 independent realizations. Given each realization for ${\bf{x}}$, obtain the secrecy outage probability by only optimizing ${\bf{w}}$ using the PGA mentioned in Algorithm 2. Finally, compute 100 results corresponding to different ${\bf{x}}$ and select the minimum value.

\item RAP+ZF: Compared to RAP+OB, the only difference is that given each realization for ${\bf{x}}$, ${\bf{w}}$ is designed by adopting the ZF-based beamforming shown in (37).

\item FPA+OB: Compared to RAP+OB, the only difference is that positions of antennas at Alice is fixed as ${\bf{x}} = {[0,0.5,1,...,(N - 1)2]^T}$.

\item FPA+ZF: Positions of antennas of Alice is fixed as ${\bf{x}} = {[0,0.5,1,...,(N - 1)2]^T}$, and ${\bf{w}}$ is designed by adopting the ZF-based beamforming shown in (37).

\item MA+MRT: Alice adopts the MRT-based beamforming and concurrently determines positions of antennas at itself using the PGA mentioned in Algorithm 2.
\end{itemize}

Fig. 8 first illustrates the secrecy outage probability w.r.t. the Rician-factor $K$, from which it is observed: i) when $K = 0$, i.e., the wiretap channel suffers from the special case of Rayleigh fading, as revealed in Insight 2, positions of antennas at Alice can be arbitrary and Alice just adopts the MRT-based beamforming to achieve the optimal performance. Hence, obviously MA+OB, RAP+OB, FPA+OB and MA+MRT can concurrently achieve the optimal solution. While due to the inappropriate beamforming, MA+ZF, RAP+ZF and FPA+ZF achieve a undesirable performance; ii) when $K > 0$, the secrecy outage probability achieved by MA+OB, RAP+OB and FPA+OB first increases and then decreases. This is due to that when $K$ increases from zero to a positive number, there will exist the sudden increase for the LoS component of the wiretap channel, and so does the averaged receiving power at eavesdroppers via the LoS paths. While when $K$ continues to increase, for the wiretap channel, the LoS component becomes dominant and the NLoS component gradually vanishes. Hence, the averaged receiving power at eavesdroppers via the NLoS paths will become smaller. At the same time, even the LoS paths are dominant, the averaged receiving power at eavesdroppers via them can also be reduced via the joint design of ${\bf{w}}$ and ${\bf{x}}$. For MA+ZF, RAP+ZF and FPA+ZF, since the averaged receiving power at eavesdroppers via the LoS paths always equals zero regardless of how $K$ is, while such power via the NLoS paths reduces as $K$ increases, the achievable secrecy outage probability of these schemes will decrease w.r.t. $K$; iii) due to the random set for positions of antennas at Alice, RAP+OB (RAP+ZF) results in a higher secrecy outage probability compared to MA+OB (MA+ZF); iv) for FPA+OB and FPA+ZF, since antennas' positions at Alice are fixed, i.e., no additional spatial DoF can be exploited, the correlation between the main channel and the wiretap channel cannot be reduced, which thus results in the larger secrecy outage probability. For instance, when $K = 7$, compared to MA+OB, FPA+OB results in the $30\%$ performance loss; v) even the ZF-based beamforming can completely eliminate the averaged receiving power at eavesdroppers via the LoS paths, such beamforming also results in a smaller power gain at Bob, especially when the AoDs ${\theta _0}$ and ${\theta _1}$ are close to each other. Hence, obviously MA+ZF (RAP+ZF) is slightly inferior to MA+OB (RAP+OB); vi) since the MRT-based beamforming just cares about the averaged receiving power at Bob and totally ignores such power at eavesdroppers, MA+MRT will achieve the worst performance when $K > 0$; vii) as $K$ increases to infinity, except for MA+MRT, predictably the secrecy outage probability achieved by the other schemes will decrease and then approach zero. The reasons have been explained in Insight 3 and are not repeated here.

\begin{figure}
\centering
\includegraphics[width=8cm]{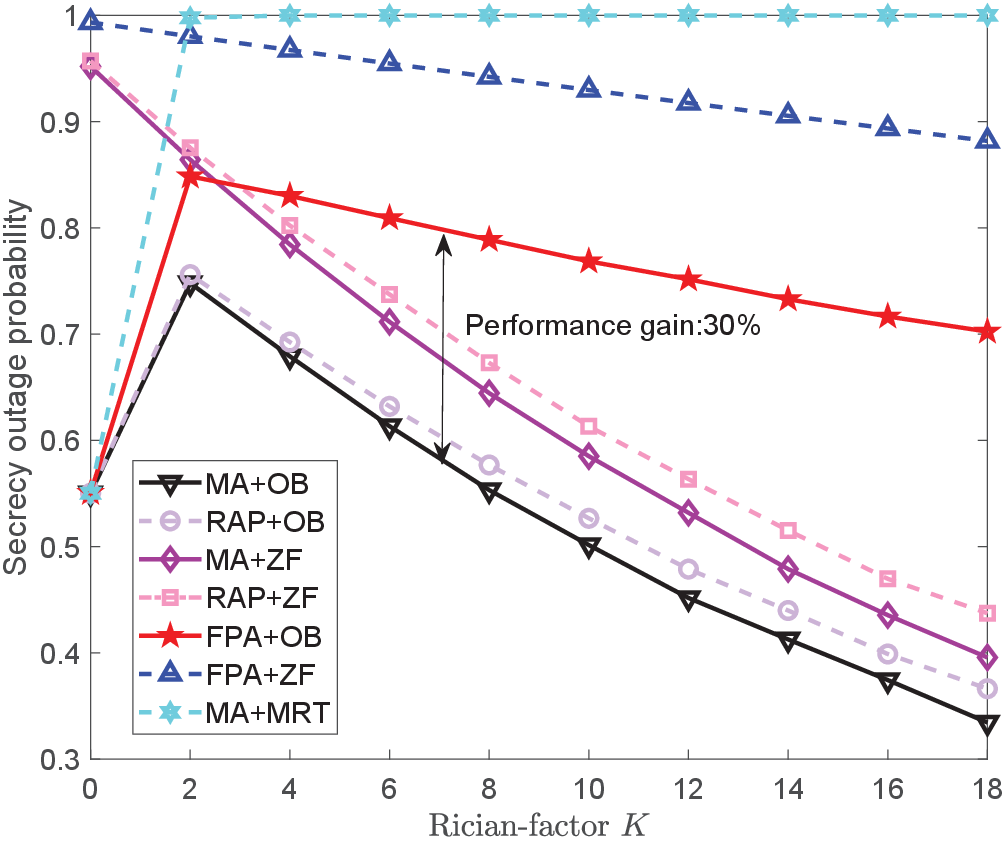}
\captionsetup{font=small}
\caption{Secrecy outage probability versus the Rician-factor $K$, where $P_a = 15$ dB, $N = 5$, $M = 1$, $L = 3$, ${\theta _0} = \frac{\pi }{4}$ and ${\theta _1} = \frac{{1.1\pi }}{4}$.} \label{fig:Fig1}
%\vspace{-5pt}
\end{figure}

\begin{figure}
\centering
\includegraphics[width=8cm]{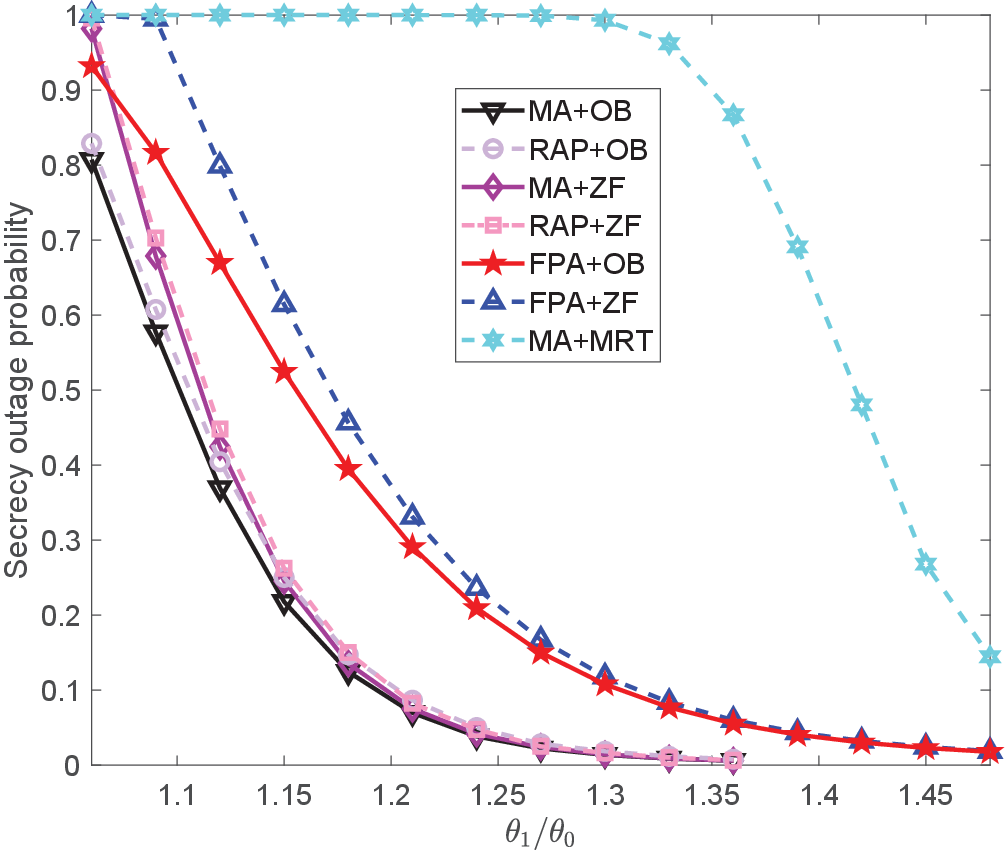}
\captionsetup{font=small}
\caption{Secrecy outage probability versus the correlation (${\theta _1}/{\theta _0}$) between the main channel and the wiretap channel, where $P_a = 15$ dB, $N = 5$, $M = 1$, $L = 3$, $K = 10$ and ${\theta _0} = \frac{\pi }{4}$.} \label{fig:Fig1}
%\vspace{-5pt}
\end{figure}

\begin{figure}
\centering
\includegraphics[width=8cm]{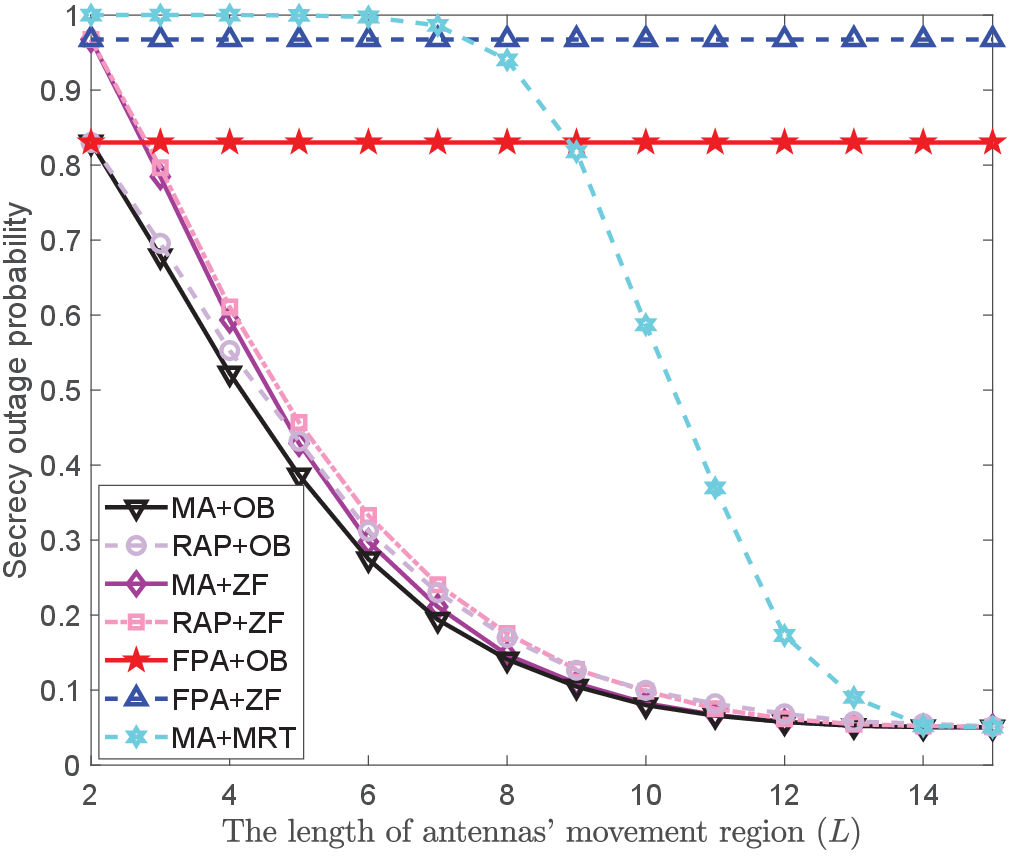}
\captionsetup{font=small}
\caption{Secrecy outage probability versus the length of antennas' movement region $L$, where $P_a = 15$ dB, $N = 5$, $M = 1$, $K = 4$, ${\theta _0} = \frac{\pi }{4}$ and ${\theta _1} = \frac{{1.1\pi }}{4}$.} \label{fig:Fig1}
%\vspace{-5pt}
\end{figure}

\begin{figure}
\centering
\includegraphics[width=8cm]{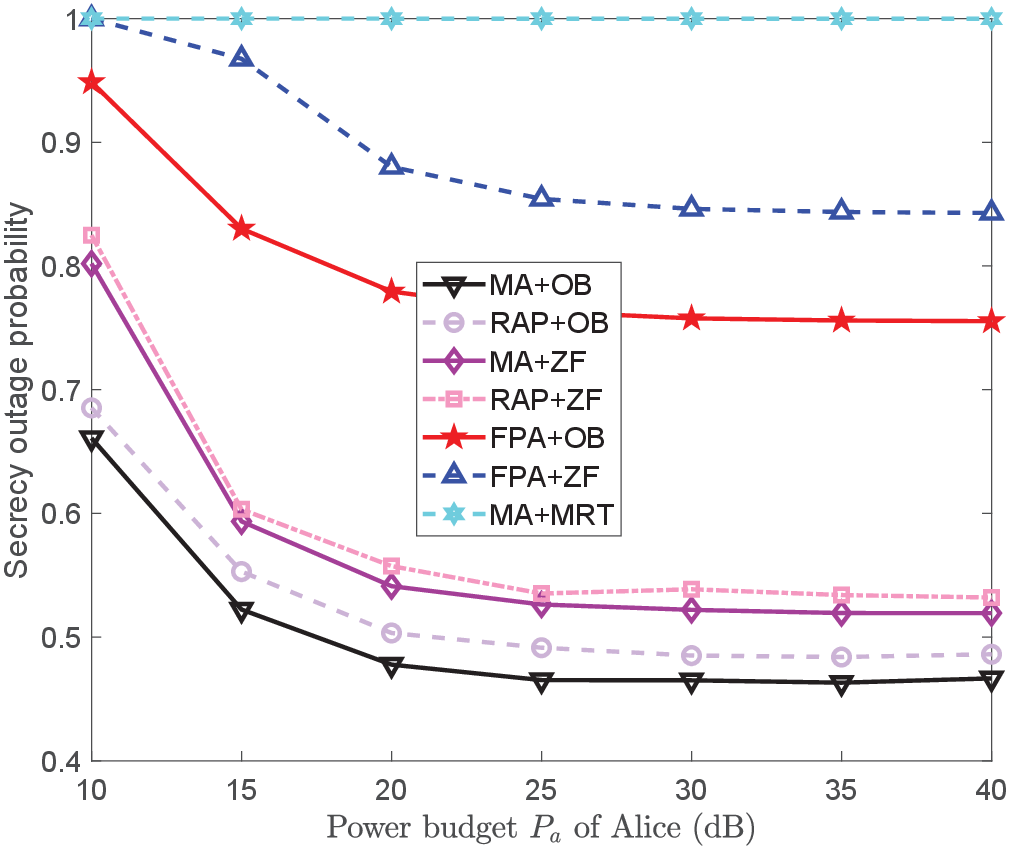}
\captionsetup{font=small}
\caption{Secrecy outage probability versus the power budget ($P_a$) of Alice, where $N = 5$, $M = 1$, $L = 4$, $K = 4$ and ${\theta _0} = \frac{\pi }{4}$, and ${\theta _1} = \frac{{1.1\pi }}{4}$.} \label{fig:Fig1}
%\vspace{-5pt}
\end{figure}

\begin{figure}
\centering
\includegraphics[width=8cm]{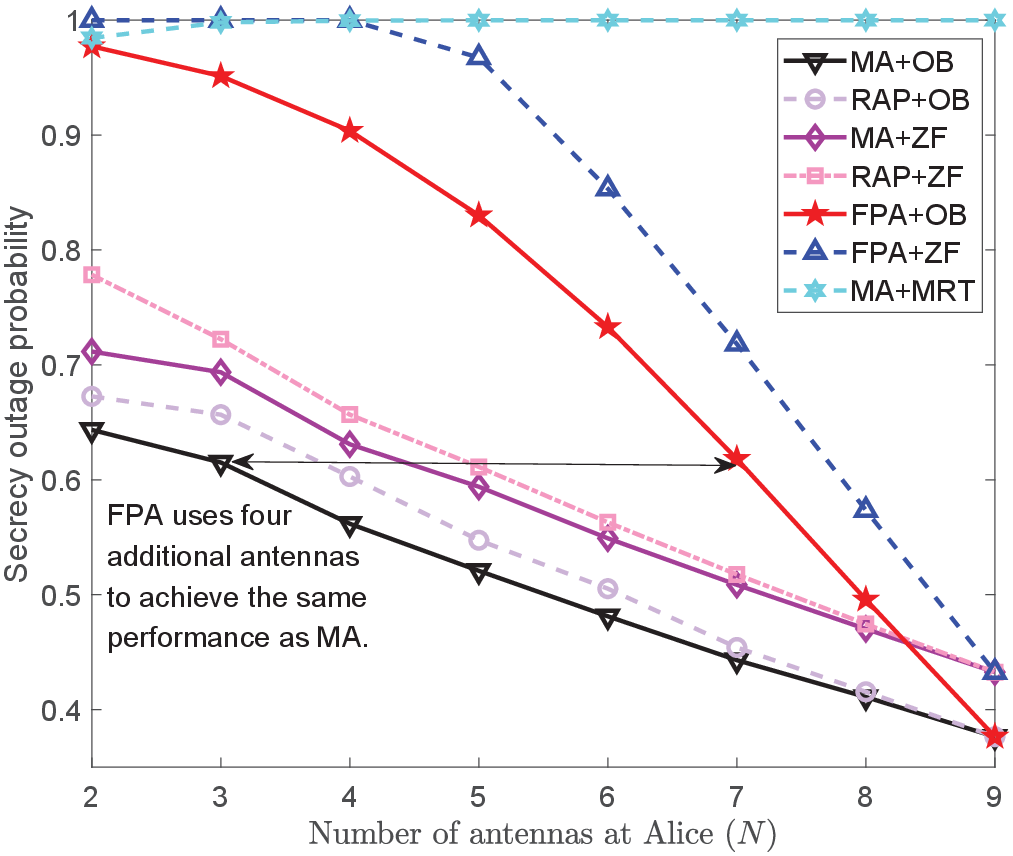}
\captionsetup{font=small}
\caption{Secrecy outage probability versus the number of antennas ($N$) at Alice, where $P_a = 15$ dB, $M = 1$, $L = 4$, $K = 4$, ${\theta _0} = \frac{\pi }{4}$, and ${\theta _1} = \frac{{1.1\pi }}{4}$.} \label{fig:Fig1}
%\vspace{-5pt}
\end{figure}

\begin{figure}
\centering
\includegraphics[width=8cm]{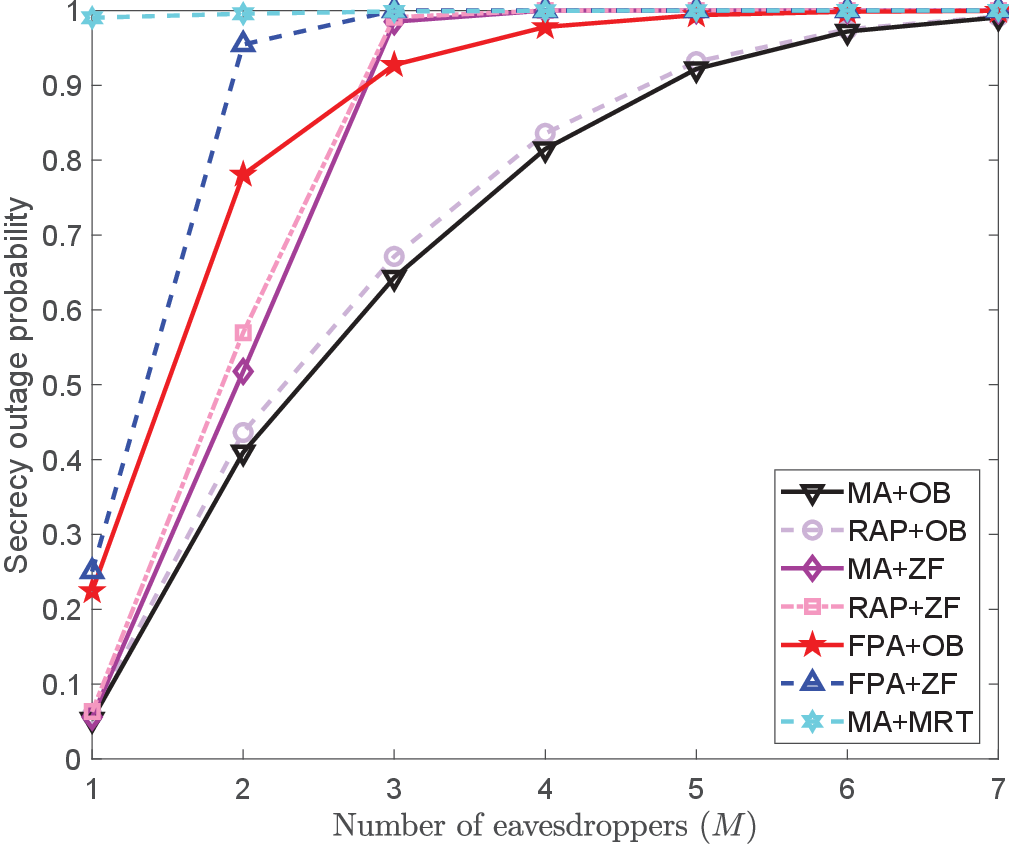}
\captionsetup{font=small}
\caption{Secrecy outage probability versus the number of eavesdroppers $M$, where $P_a = 25$ dB, $N = 8$, $L = 5.5$, $K = 4$, ${\theta _0} = \frac{\pi }{4}$, ${\theta _1} = \frac{{1.15\pi }}{4}$, ${\theta _2} = \frac{{1.35\pi }}{4}$, ${\theta _3} = \frac{{1.4\pi }}{4}$, ${\theta _4} = \frac{{1.45\pi }}{4}$, ${\theta _5} = \frac{{1.5\pi }}{4}$, ${\theta _6} = \frac{{1.55\pi }}{4}$ and ${\theta _7} = \frac{{1.6\pi }}{4}$.} \label{fig:Fig1}
%\vspace{-5pt}
\end{figure}

Fig. 9 presents the secrecy outage probability w.r.t. the correlation between the main channel and the wiretap channel, where the correlation is denoted as ${\theta _1}/{\theta _0}$. We can observe that, as ${\theta _1}/{\theta _0}$ increases, i.e., the correlation between two kinds of channels becomes smaller, the secrecy outage probability achieved by all schemes obviously reduces. Further, the performance of MA+ZF (RAP+ZF) almost approaches that of MA+OB (RAP+OB) when ${\theta _1}/{\theta _0}$ is larger than $1.25$. Therefore, it is concluded that when the correlation between the main channel and the wiretap channel reduces, considering the implementation complexity, it is enough to just adopt the ZF-based beamforming. In addition, we can similarly observe that, by exploiting the additional spatial DoF resulting from antennas' flexible movements, MA+OB (MA+ZF) achieves an excellent performance compared to FPA+OB (FPA+ZF).

Fig. 10 shows the secrecy outage probability w.r.t. the length of antennas' movement region $L$, from which we can observe that, as $L$ increases, there will exist a larger and more flexible space that can be exploited for deploying Alice's antennas to fully decrease the correlation between the main channel and the wiretap channel. Therefore, obviously the secrecy outage probability achieved by MA+OB, RAP+OB, MA+ZF, RAP+ZF and MA+MRT will become smaller w.r.t. $L$. Further, as $L$ increases, due to the smaller correlation, the ZF-based transmit beamforming will achieve the pretty good performance, indicating that MA+ZF can be adopted in a larger $L$ region. In addition, as $L$ continues to increase, the secrecy outage probability achieved by MA+OB, RAP+OB, MA+ZF, RAP+ZF and MA+MRT will converge to a constant. This phenomenon indicates that it is not necessary to expand $L$ indefinitely and only a limited span is enough to achieve the optimal performance.

Fig. 11 presents the secrecy outage probability w.r.t. the power budget ($P_a$) of Alice, from which we can observe that, as $P_a$ increases, the secrecy outage probability achieved by all schemes will first decrease and then converge to a constant. The reasons have been explained in Insight 1 and thus are omitted here.

Fig. 12 illustrates the secrecy outage probability w.r.t. the number of antennas at Alice ($N$), from which we can observe that, as $N$ increases, Alice could implement more powerful beamforming to reduce and increase the averaged receiving power at eavesdroppers and Bob, respectively. Therefore, the secrecy performance of all schemes (except for MA+MRT) becomes better. In addition, we can intuitively see that when achieving the same secrecy outage probability, e.g., 0.62, FPA+OB will employ additional four antennas (so does the number of RF chains) as compared to MA+OB. Therefore, the hardware costs can be effectively reduced by flexibly optimizing positions of antennas at Alice.

Finally, Fig. 13 shows the secrecy outage probability w.r.t. the number of eavesdroppers ($M$). We can see that as $M$ increases, more eavesdroppers can jointly decode the signal transmitted by Alice, leading to a better receiving power gain at eavesdroppers. Therefore, the secrecy performance achieved by all schemes obviously deteriorates.

  \section{Conclusion}
In this paper, secure transmission empowered by movable antennas technology is investigated, under the setup where Alice does not master the instantaneous CSI of the wiretap channel. The tight distribution of the receiving power gains at eavesdroppers is first obtained using the Laguerre series approximations, according to which the approximate expression of the secrecy outage probability is successfully derived. Afterwards, we formulate the problem of minimizing the secrecy outage probability by jointly optimizing the transmit beamforming and antennas' positions at Alice. Unfortunately, the objective includes the complex incomplete gamma function, which greatly hinders the optimization process. To tackle this, we effectively approximate the inverse of the incomplete gamma function as a simple linear model, according to which the problem can be transformed into a simple one. Then, we develop an alternating projected gradient ascent algorithm and further a projected gradient descent algorithm to tackle the problem with different implementation complexities. It is shown by simulation results the effectiveness of our proposed schemes compared to numerous benchmarks.

\normalem
\bibliographystyle{IEEEtran}
\bibliography{IEEEabrv,mybib}
\end{document}